\documentclass[%
 aip,
 amsmath,amssymb,
preprint,%
]{revtex4-1}

\usepackage{graphicx}
\usepackage{dcolumn}
\usepackage{bm}

\usepackage[utf8]{inputenc}
\usepackage[T1]{fontenc}
\usepackage{mathptmx}

\begin{document}

\preprint{AIP/123-QED}

\title[
Analysis of low-threshold optically pumped III-nitride microdisk lasers]{
Analysis of low-threshold optically pumped III-nitride microdisk lasers}

\author{Farsane Tabataba-Vakili}
\affiliation{Université Paris-Saclay, CNRS, C2N, 91120, Palaiseau, France.}
\affiliation{Univ. Grenoble Alpes, CEA, IRIG-Pheliqs, 38000 Grenoble, France.}
\author{Christelle Brimont}
\affiliation{L2C, Université de Montpellier, CNRS, 34095 Montpellier, France.}
\author{Blandine Alloing}
\affiliation{Université Côte d’Azur, CNRS, CRHEA, 06560 Valbonne, France.}%
\author{Benjamin Damilano}
\affiliation{Université Côte d’Azur, CNRS, CRHEA, 06560 Valbonne, France.}%
\author{Laetitia Doyennette}
\affiliation{L2C, Université de Montpellier, CNRS, 34095 Montpellier, France.}
\author{Thierry Guillet}
\affiliation{L2C, Université de Montpellier, CNRS, 34095 Montpellier, France.}
\author{Moustafa El Kurdi}
\affiliation{Université Paris-Saclay, CNRS, C2N, 91120, Palaiseau, France.}
\author{Sébastien Chenot}
\affiliation{Université Côte d’Azur, CNRS, CRHEA, 06560 Valbonne, France.}%
\author{Virginie Brändli}
\affiliation{Université Côte d’Azur, CNRS, CRHEA, 06560 Valbonne, France.}%
\author{Eric Frayssinet}
\affiliation{Université Côte d’Azur, CNRS, CRHEA, 06560 Valbonne, France.}%
\author{Jean-Yves Duboz}
\affiliation{Université Côte d’Azur, CNRS, CRHEA, 06560 Valbonne, France.}%
\author{Fabrice Semond}
\affiliation{Université Côte d’Azur, CNRS, CRHEA, 06560 Valbonne, France.}%
\author{Bruno Gayral}
\affiliation{Univ. Grenoble Alpes, CEA, IRIG-Pheliqs, 38000 Grenoble, France.}
\author{Philippe Boucaud}%

 \email{Philippe.Boucaud@crhea.cnrs.fr}
 
\affiliation{Université Côte d’Azur, CNRS, CRHEA, 06560 Valbonne, France.}%

\date{\today}

\begin{abstract}
Low-threshold lasing under pulsed optical pumping is demonstrated at room temperature for III-nitride microdisks with InGaN/GaN quantum wells on Si in the blue spectral range. Thresholds in the range of $18~\text{kW/cm}^2$ have been achieved along with narrow linewidths of 0.07 nm and a large peak to background dynamic of 300. We compare this threshold range with the one that can be calculated using a rate equation model. We show that thresholds in the few $\text{kW/cm}^2$ range constitute the best that can be achieved with III-nitride quantum wells at room temperature. The sensitivity of lasing on the fabrication process is also discussed.
\end{abstract}

\maketitle

Small foot-print microresonators such as microdisks are an important building block for integrated photonic circuits. The first microdisk laser was reported in 1991 by McCall et al. using the InP platform in the near-infrared (NIR) \cite{McCall1991}. To go towards the visible (VIS) and ultraviolet (UV) spectral ranges requires large band gap semiconductors and is very interesting for visible light communication \cite{Islim2017} or bio-sensing \cite{Estevez2012}. III-nitrides are the optimal candidates, as they provide a large transparency window from the UVC to the NIR and allow for the monolithic integration of active emitters in the UV-VIS spectral range. III-nitride photonic cavities have been studied for the past 15 years \cite{Butte2019}. Individual microdisk lasers have been demonstrated from the UVC to the green spectral range, mainly under pulsed optical pumping at room temperature \cite{Haberer2004, Simeonov2007, Simeonov2008, Aharonovich2013, Selles2016_1, Selles2016_2, Zhu2020}. A few reports have also been made on continuous-wave (CW) lasing in III-nitride microdisks \cite{Tamboli2007, Athanasiou2014, Athanasiou2017}. Large quality (Q) factors in the range of 7000 to 10000 have been reported \cite{Mexis2011,Rousseau2018}. Electrically injected microdisk lasers have been demonstrated in the blue and UVA under pulsed and CW pumping \cite{Kneissl2004, Feng2018, Wang2019, Wang2020}. Microdisk lasers have been monolithically integrated into photonic circuits in the blue \cite{TabatabaVakili2018, TabatabaVakili2019_2} and UVA \cite{TabatabaVakili2020} spectral range  and a scheme to combine photonic circuit and electrical injection has been proposed \cite{TabatabaVakili2019_1}.

In this letter, we will discuss our recent advances in III-nitride optically pumped microdisk lasers. We demonstrate a significantly low threshold down to $P_{th} = 18~\text{kW/cm}^2$ at room temperature under pulsed optical pumping. This low threshold comes along with a narrow linewidth and a large peak-to-background ratio. We discuss the parameters governing the threshold in III-nitride microdisks following a rate equation analysis. This includes the transparency carrier density, the cavity quality factor, the modal overlap of the active region with the confined mode, the material gain, and the carrier lifetime. We find an excellent agreement between experimental values and the calculated ones using appropriate values for these III-nitride microresonators. We will show that in order to attain the transparency carrier density $n_{tr}$ in blue InGaN quantum wells (QWs) a minimum pump power of a few kW/cm$^2$ is required. The thresholds reported in this work are thus close to the optimum for room temperature lasing with III-nitride microdisks. The rate equation analysis allows us to discuss previous values reported in the literature.

\begin{figure}[htbp]
\centering
\includegraphics[width=\linewidth]{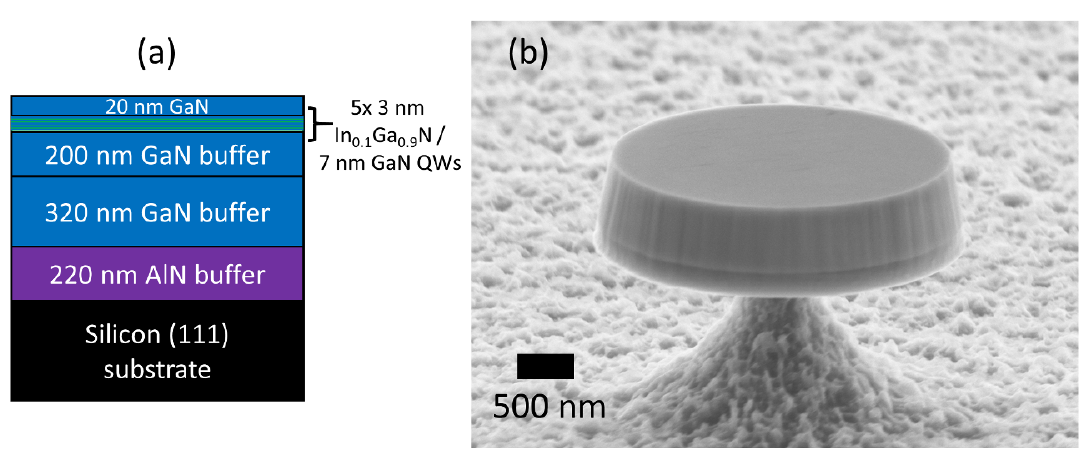}
\caption{(a) Schematic of the heterostructure of the investigated sample and (b) SEM image of a $3~ \mu \text{m}$ diameter disk.}
\label{fig:structure}
\end{figure}

The investigated sample was grown by metal organic chemical vapor deposition (MOCVD) on Si (111). In a first growth run, a template was grown consisting of 220 nm AlN and 320 nm GaN. Then in a second growth run another 200 nm of GaN were grown, followed by the active region consisting of $5\times$ 3 nm In$_{0.1}$Ga$_{0.9}$N / 7 nm GaN QWs, and a 20 nm GaN cap layer. The total thickness is 810 nm. A schematic of the heterostructure is depicted in Fig. \ref{fig:structure} (a). The threading dislocation density is estimated to be $1.2\times 10^{10} ~\text{cm}^{-2}$ from a $2\times 2$ $\mu \text{m}^2$ atomic force microscopy image (see Fig. S1 in the supplementary material), a standard value for thin III-nitride layers grown on Si. A photoluminescence (PL) measurement of the as-grown sample is shown in Fig. S2 of the supplementary material, showing QW emission centered around 428 nm. We fabricated microdisks using standard cleanroom processing. We used a plasma enhanced chemical vapor deposition (PECVD) SiO$_2$ hard mask, e-beam lithography using UV5 positive resist, and inductively coupled plasma (ICP) etching using CH$_2$F$_2$ and CF$_4$ gases for the SiO$_2$ and Cl$_2$ and BCl$_3$ gases for the III-nitride. The Si was underetched using XeF$_2$ gas. A scanning electron microscope (SEM) image of a $3~ \mu \text{m}$ diameter disk is depicted in Fig. \ref{fig:structure} (b), showing smooth slightly-inclined side-walls.

We use a standard $\mu$-PL setup with a 355 nm pulsed laser (4 ns pulse width, 7 kHz repetition rate) and a $50 \times$ microscope objective to both pump the microdisk and collect its emission from the top. A spectrometer and a  Peltier-cooled charge coupled device (CCD) are used as the detector.

\begin{figure}[htbp]
\centering
\includegraphics[width=0.6\linewidth]{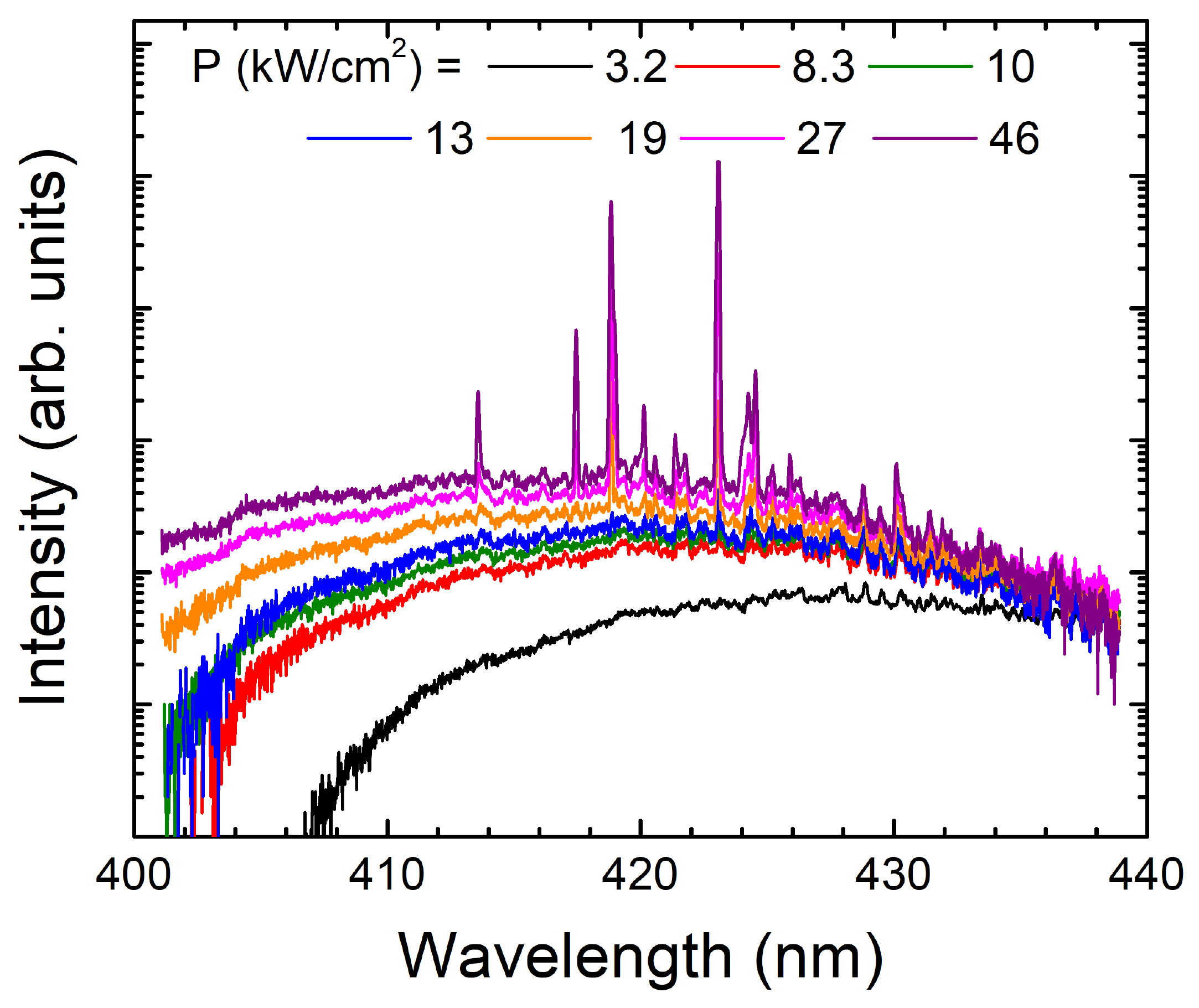}
\caption{Pump power dependent spectra of a $3~ \mu \text{m}$ diameter disk.}
\label{fig:spectra}
\end{figure}

\begin{figure}[htbp]
\centering
\includegraphics[width=\linewidth]{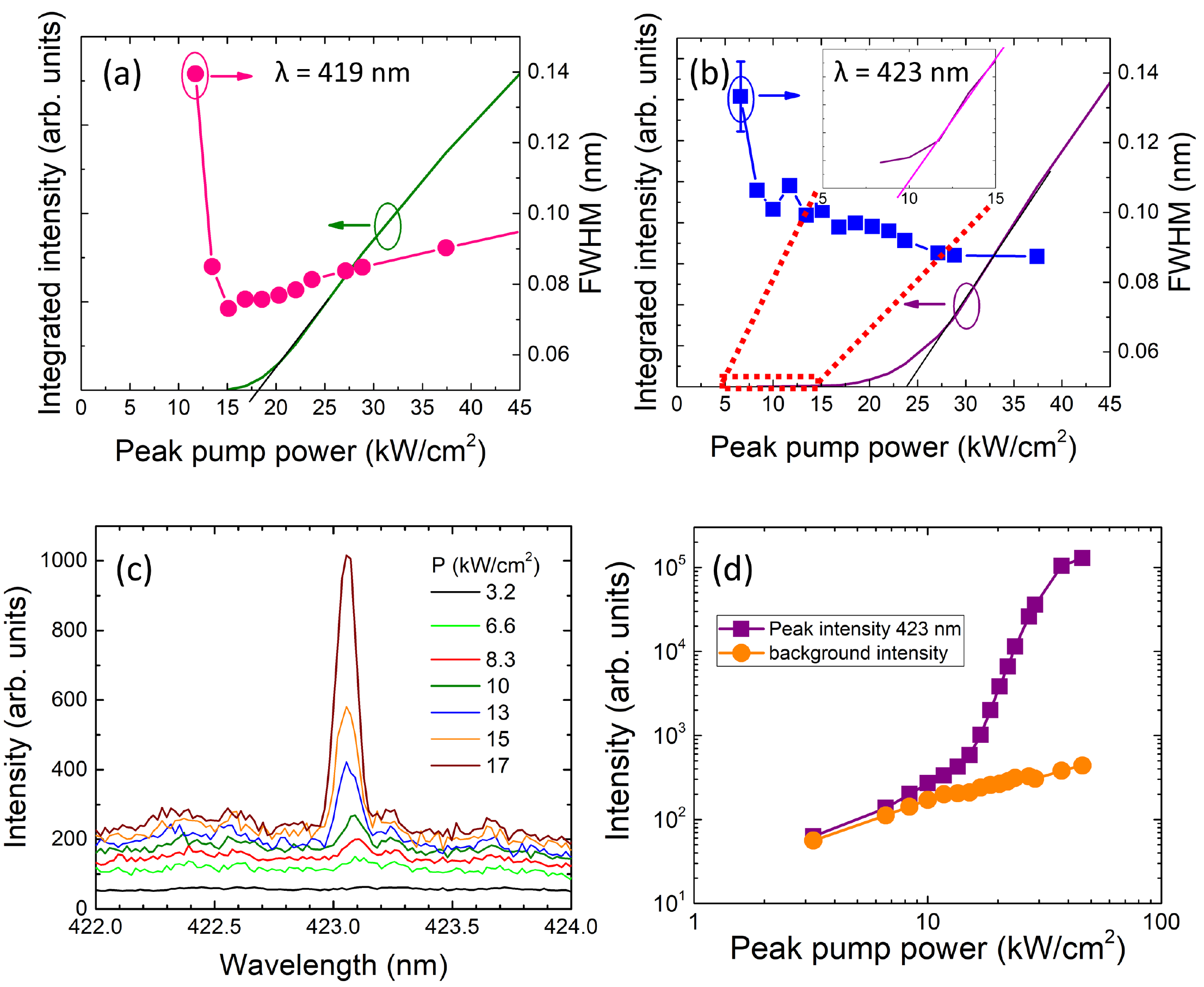}
\caption{Integrated intensity and full-width at half maximum (FWHM) of the modes (a) at 419 nm and (b) at 423 nm. Inset in (b) shows a zoom-in at low power density.  (c) Close-up of the spectra at low power density for the 423 nm mode. (d) Peak and background intensities over pump power for the mode at 423 nm.}
\label{fig:analysis}
\end{figure}

Fig. \ref{fig:spectra} shows pump power dependent spectra of a $3~ \mu \text{m}$ diameter disk. Several modes are lasing and we can see clamping of the spontaneous emission at low energy (see Fig. S3 in the supplementary material). The two strongest modes at 419 and 423 nm have very similar thresholds and are in strong competition.  Figs. \ref{fig:analysis} (a) and (b) show the integrated intensity and linewidth over peak pump power for the modes at 419 and 423 nm, respectively. We point out the limited visibility of whispering-gallery modes (WGMs) below threshold in a top-collection setup, as WGMs radiate preferentially in-plane. For the mode at 419 nm (Fig. \ref{fig:analysis} (a)), the narrowest linewidth of 0.07 nm is observed at $15~\text{kW/cm}^2$ peak power and the threshold is around $18~\text{kW/cm}^2$ (or threshold energy per pulse of $0.07 ~\text{mJ/cm}^2$). At higher power densities, a second mode appears at nearly the same wavelength (see Fig. S4 in the supplementary material), thus making the linewidth analysis complex.  For the mode at 423 nm (Fig. \ref{fig:analysis} (b)), the linewidth starts narrowing at $8~\text{kW/cm}^2$.  A zoom of the low power range of $3$-$17~\text{kW/cm}^2$, given in Fig. \ref{fig:analysis} (c), shows the onset of the mode dynamics and indicates a threshold at $10~\text{kW/cm}^2$ consistent with the linewidth narrowing. There is nonetheless mode competition with the 419 nm mode, which is discussed in more detail in the supplementary material (Fig. S5). Consequently, Fig. \ref{fig:analysis} (b))  indicates an overall threshold to be around $25~\text{kW/cm}^2$. A close-up of the mode at 423 nm at low pump powers below threshold is shown in Fig. \ref{fig:analysis} (c). A large dynamic between the peak intensity and the background spontaneous emission of $> 300$ can be discerned for the 423 nm mode in Fig. \ref{fig:analysis} (d), where we plot the peak and background intensities over pump power in a double logarithmic plot. At the last point of $45~\text{kW/cm}^2$ the CCD is saturated.

Using the rate equation model introduced by Baba and Sano, we get the following formula for the threshold power \cite{Baba2003}:

\begin{equation}
    P_{th} = \frac{E_{phot} d}{U \eta_{inj} (1-R)} \left( \frac{1-C_{sp}}{\tau_{sp}} + \frac{1}{\tau_{nr}} \right) n_{th}, \label{eq:Pth}
\end{equation}

with $E_{phot}$ the photon energy of the pump laser, $d$ the total thickness of the active medium, $U$ the ratio between the homogeneous broadening of the laser mode and the spectral broadening of the spontaneous emission \cite{Baba2003}, $\eta_{inj}$ the injection efficiency of the carriers into the QWs, $R$ the surface reflectivity, $C_{sp}$ the spontaneous emission factor, $\tau_{sp}$ the spontaneous emission lifetime, $\tau_{nr}$ the non-radiative lifetime, and $n_{th}$ the threshold carrier density, which is given by

\begin{equation}
    n_{th} = n_{tr} + \frac{n_g}{c \Gamma g \tau_c}, \label{eq:nth}
\end{equation}

where $n_{tr}$ is the transparency carrier density, $n_g$ the group index, $c$ the speed of light, $\Gamma$ the overlap of the optical mode with the QWs, $g$ the dynamic gain, and $\tau_c$ the photon lifetime. Equation \ref{eq:nth} indicates that the first requirement to obtain a net positive gain is to reach transparency. Once transparency is attained, another important factor to reach threshold is $\Gamma$. For the TE$_0$ mode, we calculate $\Gamma=0.44\%$ for our 5 QWs (see Fig. S6 in the supplementary material), which is very small and could be improved in future work. We calculate $n_g=3.44$ and $g$ is given by $g=g_0 / n_{tr}$, where $g_0$ is an empirical gain coefficient, which we estimate to be $35000~\text{cm}^{-1}$, based on Ref. \cite{Trivino2015}, where they study a fairly similar QW (3 nm, 15\% In) and which constitutes a typical value. For more information on the threshold gain, refer to the supplementary material.

The photon lifetime is given by

\begin{equation}
    \tau_c = \frac{Q \lambda}{2 \pi c}, \label{eq:tauc}
\end{equation}

with $Q$ the quality factor and $\lambda$ the peak wavelength. We calculate $Q=6000$ using the below threshold linewidth, which is in good agreement with our previous samples \cite{Selles2016_2, TabatabaVakili2019_2}, and $\lambda=420~\text{nm}$. We get $\tau_c=1.3~\text{ps}$.

 The sheet transparency carrier density is around $3\times 10^{12}~\text{cm}^{-2}$ for InGaN QWs, which gives us $n_{tr}=1\times 10^{19}~\text{cm}^{-3}$ for our 3 nm QWs \cite{Chow1996, Hangleiter1997, Trivino2015}. Using Eq. \ref{eq:nth}, we thus get $n_{th}=1.57\times 10^{19}~\text{cm}^{-3}=1.57 n_{tr}$.

We can calculate $C_{sp}$ using \cite{Baba2003}

\begin{equation}
    C_{sp} = \frac{p \Gamma_r \lambda^3}{4 \pi^2 n_{eff}^3 V_m} \frac{\lambda}{\Delta \lambda},
\end{equation}

where $p$ is the polarization anisotropy of the spontaneous emission, $\Gamma_r$ is the relative confinement factor, $n_{eff}$ is the effective refractive index, $V_m$ is the mode volume, and $\Delta \lambda$ is the homogeneous broadening of the QW emission. We calculate the mode volume using the finite-difference time-domain (FDTD) method with $V_m= \int_{V} \epsilon |\textbf{E}|^2 dV/[\epsilon |\textbf{E}|^2]_{max}$ and obtain  $0.22~\mu \text{m}^3$ for a first order radial TE$_0$ mode. Note that this weak mode volume is linked to the short wavelength investigated. Using $p=1$ (consistent with the formula for $\beta$, the spontaneous emission factor, in Ref. \cite{Coldren2012}, p. 562), $\Gamma_r = 1$ (for a cavity with many modes  \cite{Baba1997}), $n_{eff}=2.49$, and $\Delta \lambda = 15~\text{nm}$, we get $C_{sp}=2 \times 10^{-2}$. This value is still fairly small and can thus be neglected in Eq. \ref{eq:Pth}. 

Note that $C_{sp}$, the spontaneous emission factor, is in first approximation inversely proportional to the number of allowed cavity modes in the spectral broadening. This spontaneous emission factor is in the literature often presented as $\beta$  \cite{Trivino2015, Selles2016_2, Athanasiou2014}. By accounting for the homogeneous broadening in the calculation of $C_{sp}$, we assume to be in the so-called bad emitter regime, i.e. emitter linewidth larger than the cavity linewidth \cite{vanExter1996, Coldren2012}. Note that Eq. \ref{eq:Pth} is the same as in Ref. \cite{Coldren2012} (p. 252) except for the pre-factor $U$ that accounts for homogeneous broadening. 

Using the standard rate equations given in Ref. \cite{Coldren2012} (p. 249), we can fit the L-L curve of Fig. \ref{fig:analysis} (b) in the range up to $25~\text{kW/cm}^2$. We  obtain a value of $6 \times 10^{-3}$ for the spontaneous emission factor, which is within a factor of 3 of the calculated value of $C_{sp}$.
For the here presented range of microdisk parameters the equations for the laser threshold are equivalent to those for ridge waveguide lasers \cite{Morkoc2009, Coldren2012} where the spontaneous emission factor is weak.

$E_{phot}=3.49~\text{eV}$ for our pump laser and $d=15~\text{nm}$ for our 5 QWs. The reflectivity of GaN at 355 nm is 19\% \cite{Kawashima1997}. We can estimate $\tau_{sp}$ and $\tau_{nr}$ using the ABC model \cite{Scheibenzuber2011} with

\begin{equation}
    \frac{1}{\tau_{tot}} =\frac{1}{\tau_{sp}} + \frac{1}{\tau_{nr}} = A + B n +C n^2, \label{eq:tt}
\end{equation}

where the first term describes Shockley-Read-Hall recombination, the second term describes radiative recombination, and the third term describes Auger-type recombination. We use $A=4.2 \times 10^7~\text{s}^{-1}$, $B=3 \times 10^{-12}~\text{cm}^3\text{s}^{-1}$, and $C= 4.5 \times 10^{-31}~\text{cm}^6\text{s}^{-1}$ (see Refs. \cite{Scheibenzuber2011,Trivino2015}). As discussed below, there is a large spread on these values in the literature. We use the aforementioned values because they have been measured for InGaN laser diodes emitting in the same spectral range as our sample and at high carrier densities. However, the laser diodes studied by Scheibenzuber et al. \cite{Scheibenzuber2011} were grown on bulk GaN, thus having a much smaller dislocation density than our material, which has an influence on the non-radiative lifetime. Assuming a 5 times larger $A$ would increase $P_{th}$ by a factor of 2, a 10 times larger $A$ would result in a factor 3 increase in $P_{th}$. Given that our measured thresholds are quite low, the non-radiative lifetime cannot be extremely short. At threshold, we calculate $\tau_{tot}=5.1~\text{ns}$, $\tau_{sp}= 21~\text{ns}$, and $\tau_{nr}=6.6~\text{ns}$ using the ABC values of Scheibenzuber et al. \cite{Scheibenzuber2011}. It corresponds to an internal quantum efficiency of 24\% at threshold. This $\tau_{tot}$ is very close to and a bit larger than the pulse width of 4 ns, which can influence the carrier dynamics and increase the threshold. Meanwhile, a 5 times larger $A$ would result in a $\tau_{tot}$ of $2.7~\text{ns}$, which is significantly shorter than the pulse width, and a factor 2 larger $P_{th}$, which would still fit well with our experimental values. Our threshold calculation assumes a pseudo-continuous modeling. In the supplementary material, we compare the carrier lifetimes and thresholds obtained for the ABC values of Scheibenzuber et al. \cite{Scheibenzuber2011} and Espenlaub et al. \cite{Espenlaub2019} for our sample as well as the samples of Refs. \cite{Tamboli2007} and \cite{Trivino2015}.

The parameters $U$ and $\eta_{inj}$ in Eq. \ref{eq:Pth}  are not well-known. $U$ is the ratio between the homogeneous broadening and the spectral broadening. It is unlikely that all the spontaneous emission participates homogeneously to the laser mode, and $U$ is certainly less than one. Under non-resonant optical pumping, the carriers are photo-induced in the barriers and subsequently captured in the QWs. Moreover, at room temperature, a fraction of the carriers can thermally escape from the QWs. There are 5 QWs in our active structure. Considering a $U$ value of 0.4 (for a spectral broadening of 35 nm and a homogeneous broadening of 15 nm) and a large range of $\eta_{inj}$  between 0.3 and 0.7, their product is likely in the 0.1-0.3 range.

\begin{figure}[htbp]
\centering
\includegraphics[width=\linewidth]{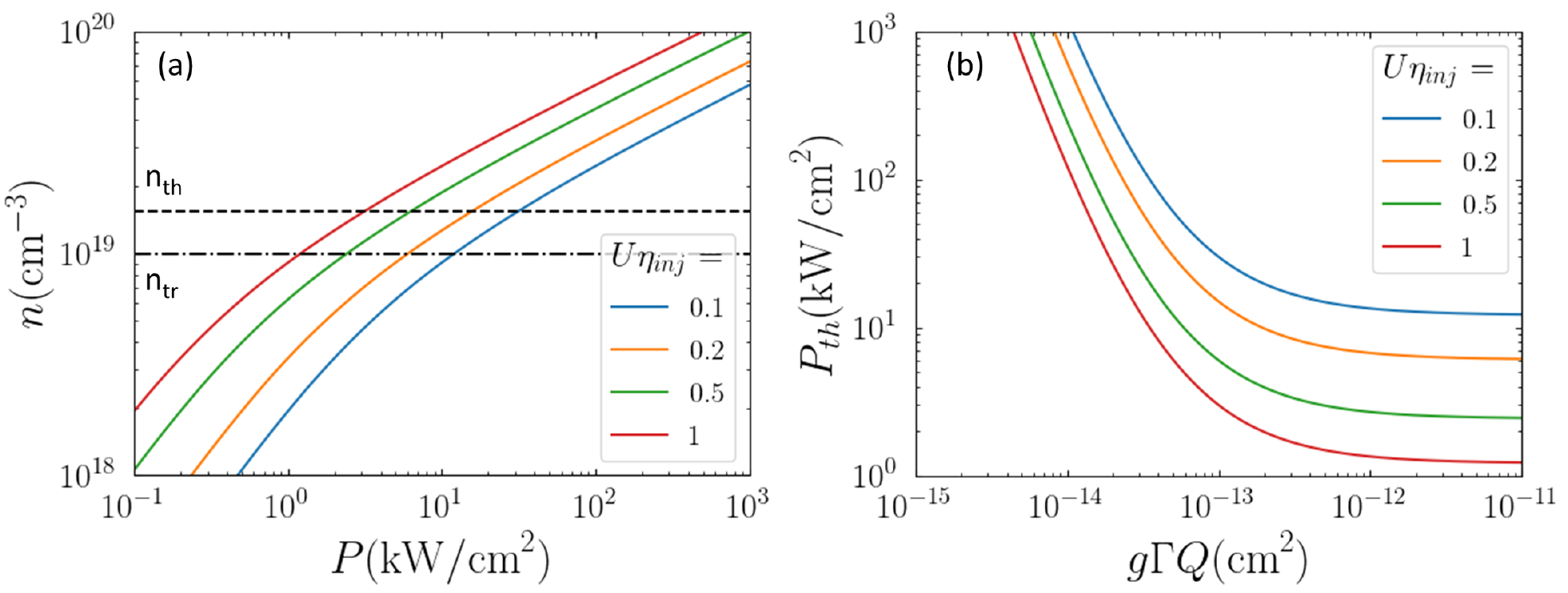}
\caption{(a) Calculation of $n$ over $P$ for different values of $U\eta_{inj}$ using Eqs. \ref{eq:Pth} and \ref{eq:tt}, and indicating the values of $n_{tr}$ and $n_{th}$. (b) Calculation of $P_{th}$ over $g\Gamma Q$ for different values of $U\eta_{inj}$ using Eqs. \ref{eq:Pth}, \ref{eq:nth}, \ref{eq:tauc}, and \ref{eq:tt}.}
\label{fig:threshold}
\end{figure}

In Fig. \ref{fig:threshold} (a), we plot $n$ over $P$ below threshold for different values of $U \eta_{inj}$ between 0.1 and 1 using Eqs. \ref{eq:Pth} (which is valid below and at threshold) and \ref{eq:tt}. The values of $n_{tr}$ and $n_{th}$ of our sample are indicated. For $U \eta_{inj}=0.18$, we get $P_{th} = 18~\text{kW/cm}^2$, which is the value we observed experimentally, indicating that the parameters considered for the III-nitride microresonators on silicon are relevant. We cannot be sure which vertical mode order is lasing. For the TE$_1$ mode, we would have a larger value of $\Gamma$, $1.6\%$, and the same $P_{th}$ value would be attained for a smaller value of $U\eta_{inj}$ of 0.1. A different vertical order would also imply a different mode volume. A $P_{th} = 10~\text{kW/cm}^2$ would correspond to a $U \eta_{inj}=0.3$. 

In Fig. \ref{fig:threshold} (b), we plot $P_{th}$ as a function of $g \Gamma Q$ using Eqs. \ref{eq:Pth}, \ref{eq:nth}, \ref{eq:tauc}, and \ref{eq:tt}. High gain, high Q and strong overlap are parameters governing the threshold. For large values of $g \Gamma Q$ (i.e. for a very high-Q cavity), $P_{th}$ goes towards $P_{tr}$, the power needed to reach transparency. We can see that even for an unrealistic $U \eta_{inj}=1$ at room temperature, $P_{tr}$ is still $1.2~\text{kW/cm}^2$. Using a single QW, as opposed to 5, can reduce the minimum power needed to reach transparency to $0.24~\text{kW/cm}^2$. We recall that for III-nitrides the relatively large threshold values as compared to other compound semiconductors are directly associated to the large density of states due to the heavy electron and holes effective masses characteristic of III-nitrides.  

In Fig. S7 in the supplementary information, we show $P_{th}$ as a function of the number of QWs.

\begin{figure}[htbp]
\centering
\includegraphics[width=0.6\linewidth]{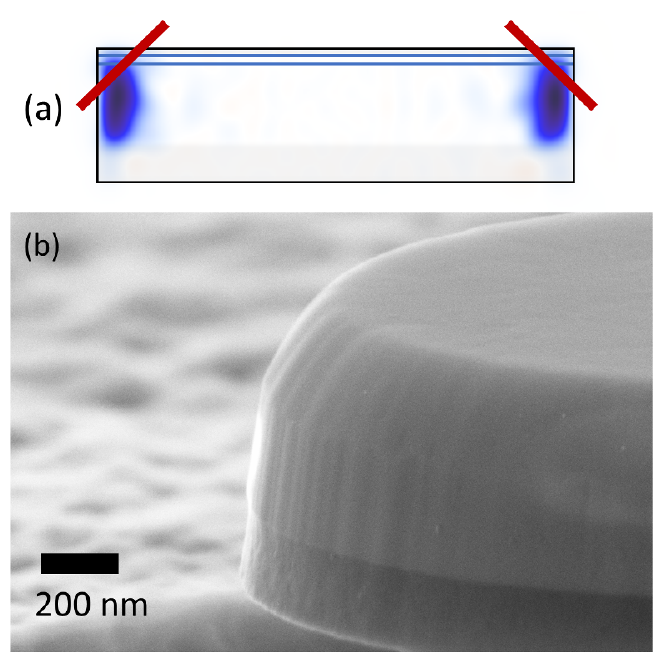}
\caption{(a) Schematic of a $3~\mu \text{m}$ diameter disk with truncated edges and depicting the $H_z$ field of a first order radial mode. (b) SEM image of a truncated microdisk.}
\label{fig:process}
\end{figure}

While low-threshold lasing has been demonstrated under electrical injection \cite{Kozaki2006}, we note that care needs to be taken when comparing such lasing thresholds with those measured under optical pumping. M. Martens \cite{Martens2018} (p. 32-33) reported for virtually identical structures a threshold of $100~\text{kW/cm}^2$ under optical pumping at 266 nm and $2~\text{kA/cm}^2$ under electrical injection. We can estimate the expected threshold current density of an equivalent structure to ours using \cite{Rosencher2004} (p. 635)

\begin{equation}
J_{th} = e d \frac{n_{th}}{\tau_{tot}},
\end{equation}

where $e$ is the elementary charge, and using $d=5 \times 3~\text{nm}$, $n_{th}=1.57\times 10^{19}~\text{cm}^{-3}$, and $\tau_{tot}=5.1~\text{ns}$ and obtain $700~\text{A/cm}^2$ for an ideal electrical injection,  whereas we have measured $18~\text{kW/cm}^2$ under optical pumping. 

Achieving a threshold below $1~\text{kW/cm}^2$ for a multi-QW (MQW) at room temperature is thus highly challenging. In the literature, a claim of lasing at threshold powers as low as $0.3~\text{kW/cm}^2$ at a similar wavelength and for a comparable active region (also 5 QWs) was reported in Ref. \cite{Tamboli2007}. They demonstrate a dynamic between peak and background of only a factor of around 8 at $9P_{th}$, which is 40 times smaller than in our case. We need to consider the large variability of ABC parameters in literature \cite{Piprek2010, Karpov2015, Espenlaub2019, David2020}. In order to achieve $P_{th}=0.3~\text{kW/cm}^2$ with the parameters of the sample reported by Tamboli et al. \cite{Tamboli2007}, a $\tau_{tot}$ of 110 ns would be required at $U \eta_{inj}=0.35$, which would require $C$ to be around one order of magnitude smaller as compared to the values of Scheibenzuber et al. \cite{Scheibenzuber2011}, along with a smaller $B$ value and a much smaller $A$ value, i.e. very long radiative lifetimes and high material quality leading to very long non-radiative lifetimes. We discuss different ABC parameters and resulting calculated thresholds in the supplementary material. This radiative lifetime improvement could be combined with an increase of the $C_{sp}$ factor by decreasing the cavity size. If a different gain medium was studied, i.e. quantum dots instead of QWs, or generally more localized emitters, the density of states would be different \cite{Arakawa1982}, and the threshold could possibly be reduced since $n_{tr}$ would be lower. Localized states are expected in InGaN QWs, but for the pump powers reported by Tamboli et al. \cite{Tamboli2007} or in this work, it is expected that they are saturated and that a 2D density of states is recovered, thus justifying the QW rate equation analysis. 

The here discussed linewidth of 0.07 nm at threshold is one of the narrowest values reported for III-nitride microdisks. The only smaller value we found, 0.033 nm, was reported by Simeonov et al. \cite{Simeonov2007,Simeonov2008} for III-nitride microdisks on sapphire with a higher threshold of $166~\text{kW/cm}^2$. As for III-nitride microdisk lasers on silicon, we have not found any narrower values.

While even a perfect resonator cannot infinitely reduce the threshold, a badly processed microdisk can easily result in no lasing being observed at all. Fig. \ref{fig:process} (a) shows a schematic of the cross-section of a $3~\mu \text{m}$ diameter disk depicting the $H_z$ field of a first order radial mode (for a perfect microdisk) and indicating truncated edges that strongly reduce the overlap of the mode with the QWs, which are at the top of the structure, not taking into account that the truncated edges would also change the mode profile a bit. In Fig. \ref{fig:process} (b) we show an SEM image of such a disk, which does not lase under pulsed optical pumping. We have also observed an order of magnitude difference in threshold for microdisks on two samples separately processed from the same wafer, highlighting the need for a high quality cavity. 

We did not observe CW lasing for our sample, which is likely due to the fact that our microdisks are fairly large ($3~\mu \text{m}$ diameter) and our CW laser is not sufficiently powerful. Meanwhile, CW lasing is always challenging for microdisk structures due to thermal management and pedestal size. A weak dependence of threshold vs. temperature is also mandatory and will be the subject of future work.

In conclusion, we have demonstrated low threshold ($18~\text{kW/cm}^2$ peak power) pulsed optically pumped lasing in III-nitride microdisks grown by MOCVD on Si with a narrow linewidth and a large peak to background dynamic of 300. We calculated the threshold power using a rate equation model and obtained values that are in good agreement with our measurements, which highlights the fact that InGaN MQWs require a minimum power in the order of kW/cm$^2$ to reach the transparency carrier density. The presented microresonators fabricated on Si are thus close to the best that can be achieved with III-nitride QWs.  

\section*{Supplementary material}
See the supplementary material for material characterization, additional lasing spectra, a 1D simulation of the mode overlap, a discussion of the threshold gain, a discussion of the dependence of the threshold on the number of QWs, as well as a discussion of the calculated lasing thresholds using different ABC values.

\begin{acknowledgments}
We thank Damir Vodenicarevic for his help with python scripts for data analysis. This work was supported by the French Agence Nationale de la Recherche (ANR) under MILAGAN convention (ANR-17-CE08-0043-02). We acknowledge support by a public grant overseen by the ANR as part of the “Investissements d’Avenir” program: Labex GANEX (Grant No. ANR-11-LABX-0014). This work was also partly supported by the RENATECH network. We acknowledge the support from the technical teams at PTA-Grenoble and Nanofab (Institut Néel).
\end{acknowledgments}

\section*{Data availability}
The data that supports the findings of this study are available within the article [and its supplementary material].

\bibliography{aipsamp}

\begin{thebibliography}{43}%
\makeatletter
\providecommand \@ifxundefined [1]{%
 \@ifx{#1\undefined}
}%
\providecommand \@ifnum [1]{%
 \ifnum #1\expandafter \@firstoftwo
 \else \expandafter \@secondoftwo
 \fi
}%
\providecommand \@ifx [1]{%
 \ifx #1\expandafter \@firstoftwo
 \else \expandafter \@secondoftwo
 \fi
}%
\providecommand \natexlab [1]{#1}%
\providecommand \enquote  [1]{``#1''}%
\providecommand \bibnamefont  [1]{#1}%
\providecommand \bibfnamefont [1]{#1}%
\providecommand \citenamefont [1]{#1}%
\providecommand \href@noop [0]{\@secondoftwo}%
\providecommand \href [0]{\begingroup \@sanitize@url \@href}%
\providecommand \@href[1]{\@@startlink{#1}\@@href}%
\providecommand \@@href[1]{\endgroup#1\@@endlink}%
\providecommand \@sanitize@url [0]{\catcode `\\12\catcode `\$12\catcode
  `\&12\catcode `\#12\catcode `\^12\catcode `\_12\catcode `\%12\relax}%
\providecommand \@@startlink[1]{}%
\providecommand \@@endlink[0]{}%
\providecommand \url  [0]{\begingroup\@sanitize@url \@url }%
\providecommand \@url [1]{\endgroup\@href {#1}{\urlprefix }}%
\providecommand \urlprefix  [0]{URL }%
\providecommand \Eprint [0]{\href }%
\providecommand \doibase [0]{http://dx.doi.org/}%
\providecommand \selectlanguage [0]{\@gobble}%
\providecommand \bibinfo  [0]{\@secondoftwo}%
\providecommand \bibfield  [0]{\@secondoftwo}%
\providecommand \translation [1]{[#1]}%
\providecommand \BibitemOpen [0]{}%
\providecommand \bibitemStop [0]{}%
\providecommand \bibitemNoStop [0]{.\EOS\space}%
\providecommand \EOS [0]{\spacefactor3000\relax}%
\providecommand \BibitemShut  [1]{\csname bibitem#1\endcsname}%
\let\auto@bib@innerbib\@empty
\bibitem [{\citenamefont {McCall}\ \emph {et~al.}(1992)\citenamefont {McCall},
  \citenamefont {Levi}, \citenamefont {Slusher}, \citenamefont {Pearton},\ and\
  \citenamefont {Logan}}]{McCall1991}%
  \BibitemOpen
  \bibfield  {author} {\bibinfo {author} {\bibfnamefont {S.~L.}\ \bibnamefont
  {McCall}}, \bibinfo {author} {\bibfnamefont {A.~F.~J.}\ \bibnamefont {Levi}},
  \bibinfo {author} {\bibfnamefont {R.~E.}\ \bibnamefont {Slusher}}, \bibinfo
  {author} {\bibfnamefont {S.~J.}\ \bibnamefont {Pearton}}, \ and\ \bibinfo
  {author} {\bibfnamefont {R.~A.}\ \bibnamefont {Logan}},\ }\bibfield  {title}
  {\enquote {\bibinfo {title} {Whispering‐gallery mode microdisk lasers},}\
  }\href {\doibase 10.1063/1.106688@apl.2019.APLCLASS2019.issue-1} {\bibfield
  {journal} {\bibinfo  {journal} {Appl. Phys. Lett.}\ }\textbf {\bibinfo
  {volume} {60}},\ \bibinfo {pages} {289--291} (\bibinfo {year}
  {1992})}\BibitemShut {NoStop}%
\bibitem [{\citenamefont {Islim}\ \emph {et~al.}(2017)\citenamefont {Islim},
  \citenamefont {Ferreira}, \citenamefont {He}, \citenamefont {Xie},
  \citenamefont {Videv}, \citenamefont {Viola}, \citenamefont {Watson},
  \citenamefont {Bamiedakis}, \citenamefont {Penty}, \citenamefont {White},
  \citenamefont {Kelly}, \citenamefont {Gu}, \citenamefont {Haas},\ and\
  \citenamefont {Dawson}}]{Islim2017}%
  \BibitemOpen
  \bibfield  {author} {\bibinfo {author} {\bibfnamefont {M.~S.}\ \bibnamefont
  {Islim}}, \bibinfo {author} {\bibfnamefont {R.~X.}\ \bibnamefont {Ferreira}},
  \bibinfo {author} {\bibfnamefont {X.}~\bibnamefont {He}}, \bibinfo {author}
  {\bibfnamefont {E.}~\bibnamefont {Xie}}, \bibinfo {author} {\bibfnamefont
  {S.}~\bibnamefont {Videv}}, \bibinfo {author} {\bibfnamefont
  {S.}~\bibnamefont {Viola}}, \bibinfo {author} {\bibfnamefont
  {S.}~\bibnamefont {Watson}}, \bibinfo {author} {\bibfnamefont
  {N.}~\bibnamefont {Bamiedakis}}, \bibinfo {author} {\bibfnamefont {R.~V.}\
  \bibnamefont {Penty}}, \bibinfo {author} {\bibfnamefont {I.~H.}\ \bibnamefont
  {White}}, \bibinfo {author} {\bibfnamefont {A.~E.}\ \bibnamefont {Kelly}},
  \bibinfo {author} {\bibfnamefont {E.}~\bibnamefont {Gu}}, \bibinfo {author}
  {\bibfnamefont {H.}~\bibnamefont {Haas}}, \ and\ \bibinfo {author}
  {\bibfnamefont {M.~D.}\ \bibnamefont {Dawson}},\ }\bibfield  {title}
  {\enquote {\bibinfo {title} {Towards 10 {G}b/s orthogonal frequency division
  multiplexing-based visible light communication using a {G}a{N} violet
  micro-{LED}},}\ }\href {\doibase 10.1364/PRJ.5.000A35} {\bibfield  {journal}
  {\bibinfo  {journal} {Photonics Res.}\ }\textbf {\bibinfo {volume} {5}},\
  \bibinfo {pages} {A35--A43} (\bibinfo {year} {2017})}\BibitemShut {NoStop}%
\bibitem [{\citenamefont {Estevez}, \citenamefont {Alvarez},\ and\
  \citenamefont {Lechuga}(2012)}]{Estevez2012}%
  \BibitemOpen
  \bibfield  {author} {\bibinfo {author} {\bibfnamefont {M.~C.}\ \bibnamefont
  {Estevez}}, \bibinfo {author} {\bibfnamefont {M.}~\bibnamefont {Alvarez}}, \
  and\ \bibinfo {author} {\bibfnamefont {L.~M.}\ \bibnamefont {Lechuga}},\
  }\bibfield  {title} {\enquote {\bibinfo {title} {Integrated optical devices
  for lab-on-a-chip biosensing applications},}\ }\href@noop {} {\bibfield
  {journal} {\bibinfo  {journal} {Laser \& Photonics Reviews}\ }\textbf
  {\bibinfo {volume} {6}},\ \bibinfo {pages} {463--487} (\bibinfo {year}
  {2012})}\BibitemShut {NoStop}%
\bibitem [{\citenamefont {Butt{\'e}}\ and\ \citenamefont
  {Grandjean}(2020)}]{Butte2019}%
  \BibitemOpen
  \bibfield  {author} {\bibinfo {author} {\bibfnamefont {R.}~\bibnamefont
  {Butt{\'e}}}\ and\ \bibinfo {author} {\bibfnamefont {N.}~\bibnamefont
  {Grandjean}},\ }\bibfield  {title} {\enquote {\bibinfo {title} {{III}-nitride
  photonic cavities},}\ }\href@noop {} {\bibfield  {journal} {\bibinfo
  {journal} {Nanophotonics}\ }\textbf {\bibinfo {volume} {9}},\ \bibinfo
  {pages} {569--598} (\bibinfo {year} {2020})}\BibitemShut {NoStop}%
\bibitem [{\citenamefont {Haberer}\ \emph {et~al.}(2004)\citenamefont
  {Haberer}, \citenamefont {Sharma}, \citenamefont {Meier}, \citenamefont
  {Stonas}, \citenamefont {Nakamura}, \citenamefont {DenBaars},\ and\
  \citenamefont {Hu}}]{Haberer2004}%
  \BibitemOpen
  \bibfield  {author} {\bibinfo {author} {\bibfnamefont {E.~D.}\ \bibnamefont
  {Haberer}}, \bibinfo {author} {\bibfnamefont {R.}~\bibnamefont {Sharma}},
  \bibinfo {author} {\bibfnamefont {C.}~\bibnamefont {Meier}}, \bibinfo
  {author} {\bibfnamefont {A.~R.}\ \bibnamefont {Stonas}}, \bibinfo {author}
  {\bibfnamefont {S.}~\bibnamefont {Nakamura}}, \bibinfo {author}
  {\bibfnamefont {S.~P.}\ \bibnamefont {DenBaars}}, \ and\ \bibinfo {author}
  {\bibfnamefont {E.~L.}\ \bibnamefont {Hu}},\ }\bibfield  {title} {\enquote
  {\bibinfo {title} {Free-standing, optically pumped, {G}a{N}/{I}n{G}a{N}
  microdisk lasers fabricated by photoelectrochemical etching},}\ }\href
  {\doibase 10.1063/1.1829167} {\bibfield  {journal} {\bibinfo  {journal}
  {Appl. Phys. Lett.}\ }\textbf {\bibinfo {volume} {84}},\ \bibinfo {pages}
  {5179--5181} (\bibinfo {year} {2004})}\BibitemShut {NoStop}%
\bibitem [{\citenamefont {Simeonov}\ \emph {et~al.}(2007)\citenamefont
  {Simeonov}, \citenamefont {Feltin}, \citenamefont {B\"uhlmann}, \citenamefont
  {Zhu}, \citenamefont {Castiglia}, \citenamefont {Mosca}, \citenamefont
  {Carlin}, \citenamefont {Butt\'e},\ and\ \citenamefont
  {Grandjean}}]{Simeonov2007}%
  \BibitemOpen
  \bibfield  {author} {\bibinfo {author} {\bibfnamefont {D.}~\bibnamefont
  {Simeonov}}, \bibinfo {author} {\bibfnamefont {E.}~\bibnamefont {Feltin}},
  \bibinfo {author} {\bibfnamefont {H.-J.}\ \bibnamefont {B\"uhlmann}},
  \bibinfo {author} {\bibfnamefont {T.}~\bibnamefont {Zhu}}, \bibinfo {author}
  {\bibfnamefont {A.}~\bibnamefont {Castiglia}}, \bibinfo {author}
  {\bibfnamefont {M.}~\bibnamefont {Mosca}}, \bibinfo {author} {\bibfnamefont
  {J.-F.}\ \bibnamefont {Carlin}}, \bibinfo {author} {\bibfnamefont
  {R.}~\bibnamefont {Butt\'e}}, \ and\ \bibinfo {author} {\bibfnamefont
  {N.}~\bibnamefont {Grandjean}},\ }\bibfield  {title} {\enquote {\bibinfo
  {title} {Blue lasing at room temperature in high quality factor
  {G}a{N}/{A}l{I}n{N} microdisks with {I}n{G}a{N} quantum wells},}\ }\href
  {\doibase 10.1063/1.2460234} {\bibfield  {journal} {\bibinfo  {journal}
  {Appl. Phys. Lett.}\ }\textbf {\bibinfo {volume} {90}},\ \bibinfo {pages}
  {061106} (\bibinfo {year} {2007})}\BibitemShut {NoStop}%
\bibitem [{\citenamefont {Simeonov}\ \emph {et~al.}(2008)\citenamefont
  {Simeonov}, \citenamefont {Feltin}, \citenamefont {Altoukhov}, \citenamefont
  {Castiglia}, \citenamefont {Carlin}, \citenamefont {Butt\'e},\ and\
  \citenamefont {Grandjean}}]{Simeonov2008}%
  \BibitemOpen
  \bibfield  {author} {\bibinfo {author} {\bibfnamefont {D.}~\bibnamefont
  {Simeonov}}, \bibinfo {author} {\bibfnamefont {E.}~\bibnamefont {Feltin}},
  \bibinfo {author} {\bibfnamefont {A.}~\bibnamefont {Altoukhov}}, \bibinfo
  {author} {\bibfnamefont {A.}~\bibnamefont {Castiglia}}, \bibinfo {author}
  {\bibfnamefont {J.-F.}\ \bibnamefont {Carlin}}, \bibinfo {author}
  {\bibfnamefont {R.}~\bibnamefont {Butt\'e}}, \ and\ \bibinfo {author}
  {\bibfnamefont {N.}~\bibnamefont {Grandjean}},\ }\bibfield  {title} {\enquote
  {\bibinfo {title} {High quality nitride based microdisks obtained via
  selective wet etching of {A}l{I}n{N} sacrificial layers},}\ }\href {\doibase
  10.1063/1.2917452} {\bibfield  {journal} {\bibinfo  {journal} {Appl. Phys.
  Lett.}\ }\textbf {\bibinfo {volume} {92}},\ \bibinfo {pages} {171102}
  (\bibinfo {year} {2008})}\BibitemShut {NoStop}%
\bibitem [{\citenamefont {Aharonovich}\ \emph {et~al.}(2013)\citenamefont
  {Aharonovich}, \citenamefont {Woolf}, \citenamefont {Russell}, \citenamefont
  {Zhu}, \citenamefont {Niu}, \citenamefont {Kappers}, \citenamefont {Oliver},\
  and\ \citenamefont {Hu}}]{Aharonovich2013}%
  \BibitemOpen
  \bibfield  {author} {\bibinfo {author} {\bibfnamefont {I.}~\bibnamefont
  {Aharonovich}}, \bibinfo {author} {\bibfnamefont {A.}~\bibnamefont {Woolf}},
  \bibinfo {author} {\bibfnamefont {K.~J.}\ \bibnamefont {Russell}}, \bibinfo
  {author} {\bibfnamefont {T.}~\bibnamefont {Zhu}}, \bibinfo {author}
  {\bibfnamefont {N.}~\bibnamefont {Niu}}, \bibinfo {author} {\bibfnamefont
  {M.~J.}\ \bibnamefont {Kappers}}, \bibinfo {author} {\bibfnamefont {R.~A.}\
  \bibnamefont {Oliver}}, \ and\ \bibinfo {author} {\bibfnamefont {E.~L.}\
  \bibnamefont {Hu}},\ }\bibfield  {title} {\enquote {\bibinfo {title} {Low
  threshold, room-temperature microdisk lasers in the blue spectral range},}\
  }\href {\doibase 10.1063/1.4813471} {\bibfield  {journal} {\bibinfo
  {journal} {Appl. Phys. Lett.}\ }\textbf {\bibinfo {volume} {103}},\ \bibinfo
  {pages} {021112} (\bibinfo {year} {2013})}\BibitemShut {NoStop}%
\bibitem [{\citenamefont {Sell\'es}\ \emph
  {et~al.}(2016{\natexlab{a}})\citenamefont {Sell\'es}, \citenamefont
  {Brimont}, \citenamefont {Cassabois}, \citenamefont {Valvin}, \citenamefont
  {Guillet}, \citenamefont {Roland}, \citenamefont {Zeng}, \citenamefont
  {Checoury}, \citenamefont {Boucaud}, \citenamefont {Mexis}, \citenamefont
  {Semond},\ and\ \citenamefont {Gayral}}]{Selles2016_1}%
  \BibitemOpen
  \bibfield  {author} {\bibinfo {author} {\bibfnamefont {J.}~\bibnamefont
  {Sell\'es}}, \bibinfo {author} {\bibfnamefont {C.}~\bibnamefont {Brimont}},
  \bibinfo {author} {\bibfnamefont {G.}~\bibnamefont {Cassabois}}, \bibinfo
  {author} {\bibfnamefont {P.}~\bibnamefont {Valvin}}, \bibinfo {author}
  {\bibfnamefont {T.}~\bibnamefont {Guillet}}, \bibinfo {author} {\bibfnamefont
  {I.}~\bibnamefont {Roland}}, \bibinfo {author} {\bibfnamefont
  {Y.}~\bibnamefont {Zeng}}, \bibinfo {author} {\bibfnamefont {X.}~\bibnamefont
  {Checoury}}, \bibinfo {author} {\bibfnamefont {P.}~\bibnamefont {Boucaud}},
  \bibinfo {author} {\bibfnamefont {M.}~\bibnamefont {Mexis}}, \bibinfo
  {author} {\bibfnamefont {F.}~\bibnamefont {Semond}}, \ and\ \bibinfo {author}
  {\bibfnamefont {B.}~\bibnamefont {Gayral}},\ }\bibfield  {title} {\enquote
  {\bibinfo {title} {Deep-{UV} nitride-on-silicon microdisk lasers},}\ }\href
  {\doibase 10.1038/srep21650} {\bibfield  {journal} {\bibinfo  {journal} {Sci.
  Rep.}\ }\textbf {\bibinfo {volume} {6}},\ \bibinfo {pages} {21650} (\bibinfo
  {year} {2016}{\natexlab{a}})}\BibitemShut {NoStop}%
\bibitem [{\citenamefont {Sell\'es}\ \emph
  {et~al.}(2016{\natexlab{b}})\citenamefont {Sell\'es}, \citenamefont {Crepel},
  \citenamefont {Roland}, \citenamefont {El$\thinspace$Kurdi}, \citenamefont
  {Checoury}, \citenamefont {Boucaud}, \citenamefont {Mexis}, \citenamefont
  {Leroux}, \citenamefont {Damilano}, \citenamefont {Rennesson}, \citenamefont
  {Semond}, \citenamefont {Gayral}, \citenamefont {Brimont},\ and\
  \citenamefont {Guillet}}]{Selles2016_2}%
  \BibitemOpen
  \bibfield  {author} {\bibinfo {author} {\bibfnamefont {J.}~\bibnamefont
  {Sell\'es}}, \bibinfo {author} {\bibfnamefont {V.}~\bibnamefont {Crepel}},
  \bibinfo {author} {\bibfnamefont {I.}~\bibnamefont {Roland}}, \bibinfo
  {author} {\bibfnamefont {M.}~\bibnamefont {El$\thinspace$Kurdi}}, \bibinfo
  {author} {\bibfnamefont {X.}~\bibnamefont {Checoury}}, \bibinfo {author}
  {\bibfnamefont {P.}~\bibnamefont {Boucaud}}, \bibinfo {author} {\bibfnamefont
  {M.}~\bibnamefont {Mexis}}, \bibinfo {author} {\bibfnamefont
  {M.}~\bibnamefont {Leroux}}, \bibinfo {author} {\bibfnamefont
  {B.}~\bibnamefont {Damilano}}, \bibinfo {author} {\bibfnamefont
  {S.}~\bibnamefont {Rennesson}}, \bibinfo {author} {\bibfnamefont
  {F.}~\bibnamefont {Semond}}, \bibinfo {author} {\bibfnamefont
  {B.}~\bibnamefont {Gayral}}, \bibinfo {author} {\bibfnamefont
  {C.}~\bibnamefont {Brimont}}, \ and\ \bibinfo {author} {\bibfnamefont
  {T.}~\bibnamefont {Guillet}},\ }\bibfield  {title} {\enquote {\bibinfo
  {title} {{III}-nitride-on-silicon microdisk lasers from the blue to the deep
  ultra-violet},}\ }\href {\doibase 10.1063/1.4971357} {\bibfield  {journal}
  {\bibinfo  {journal} {Appl. Phys. Lett.}\ }\textbf {\bibinfo {volume}
  {109}},\ \bibinfo {pages} {231101} (\bibinfo {year}
  {2016}{\natexlab{b}})}\BibitemShut {NoStop}%
\bibitem [{\citenamefont {Zhu}\ \emph {et~al.}(2020)\citenamefont {Zhu},
  \citenamefont {Li}, \citenamefont {Zhang}, \citenamefont {Li}, \citenamefont
  {Dai}, \citenamefont {Cui}, \citenamefont {Song}, \citenamefont {Xu},\ and\
  \citenamefont {Wang}}]{Zhu2020}%
  \BibitemOpen
  \bibfield  {author} {\bibinfo {author} {\bibfnamefont {G.}~\bibnamefont
  {Zhu}}, \bibinfo {author} {\bibfnamefont {J.}~\bibnamefont {Li}}, \bibinfo
  {author} {\bibfnamefont {N.}~\bibnamefont {Zhang}}, \bibinfo {author}
  {\bibfnamefont {X.}~\bibnamefont {Li}}, \bibinfo {author} {\bibfnamefont
  {J.}~\bibnamefont {Dai}}, \bibinfo {author} {\bibfnamefont {Q.}~\bibnamefont
  {Cui}}, \bibinfo {author} {\bibfnamefont {Q.}~\bibnamefont {Song}}, \bibinfo
  {author} {\bibfnamefont {C.}~\bibnamefont {Xu}}, \ and\ \bibinfo {author}
  {\bibfnamefont {Y.}~\bibnamefont {Wang}},\ }\bibfield  {title} {\enquote
  {\bibinfo {title} {Whispering-gallery mode lasing in a floating {G}a{N}
  microdisk with a vertical slit},}\ }\href@noop {} {\bibfield  {journal}
  {\bibinfo  {journal} {Sci. Rep.}\ }\textbf {\bibinfo {volume} {10}},\
  \bibinfo {pages} {253} (\bibinfo {year} {2020})}\BibitemShut {NoStop}%
\bibitem [{\citenamefont {Tamboli}\ \emph {et~al.}(2007)\citenamefont
  {Tamboli}, \citenamefont {Haberer}, \citenamefont {Sharma}, \citenamefont
  {Lee}, \citenamefont {Nakamura},\ and\ \citenamefont {Hu}}]{Tamboli2007}%
  \BibitemOpen
  \bibfield  {author} {\bibinfo {author} {\bibfnamefont {A.~C.}\ \bibnamefont
  {Tamboli}}, \bibinfo {author} {\bibfnamefont {E.~D.}\ \bibnamefont
  {Haberer}}, \bibinfo {author} {\bibfnamefont {R.}~\bibnamefont {Sharma}},
  \bibinfo {author} {\bibfnamefont {K.~H.}\ \bibnamefont {Lee}}, \bibinfo
  {author} {\bibfnamefont {S.}~\bibnamefont {Nakamura}}, \ and\ \bibinfo
  {author} {\bibfnamefont {E.~L.}\ \bibnamefont {Hu}},\ }\bibfield  {title}
  {\enquote {\bibinfo {title} {Room-temperature continuous-wave lasing in
  {G}a{N}/{I}n{G}a{N} microdisks},}\ }\href {\doibase 10.1038/nphoton.2006.52}
  {\bibfield  {journal} {\bibinfo  {journal} {Nature Photonics}\ }\textbf
  {\bibinfo {volume} {1}},\ \bibinfo {pages} {61--64} (\bibinfo {year}
  {2007})}\BibitemShut {NoStop}%
\bibitem [{\citenamefont {Athanasiou}\ \emph {et~al.}(2014)\citenamefont
  {Athanasiou}, \citenamefont {Smith}, \citenamefont {Liu},\ and\ \citenamefont
  {Wang}}]{Athanasiou2014}%
  \BibitemOpen
  \bibfield  {author} {\bibinfo {author} {\bibfnamefont {M.}~\bibnamefont
  {Athanasiou}}, \bibinfo {author} {\bibfnamefont {R.}~\bibnamefont {Smith}},
  \bibinfo {author} {\bibfnamefont {B.}~\bibnamefont {Liu}}, \ and\ \bibinfo
  {author} {\bibfnamefont {T.}~\bibnamefont {Wang}},\ }\bibfield  {title}
  {\enquote {\bibinfo {title} {Room temperature continuous-wave green lasing
  from an {I}n{G}a{N} microdisk on silicon},}\ }\href {\doibase
  10.1038/srep07250} {\bibfield  {journal} {\bibinfo  {journal} {Sci. Rep.}\
  }\textbf {\bibinfo {volume} {4}},\ \bibinfo {pages} {7250} (\bibinfo {year}
  {2014})}\BibitemShut {NoStop}%
\bibitem [{\citenamefont {Athanasiou}\ \emph {et~al.}(2017)\citenamefont
  {Athanasiou}, \citenamefont {Smith}, \citenamefont {Pugh}, \citenamefont
  {Gong}, \citenamefont {Cryan},\ and\ \citenamefont {Wang}}]{Athanasiou2017}%
  \BibitemOpen
  \bibfield  {author} {\bibinfo {author} {\bibfnamefont {M.}~\bibnamefont
  {Athanasiou}}, \bibinfo {author} {\bibfnamefont {R.~M.}\ \bibnamefont
  {Smith}}, \bibinfo {author} {\bibfnamefont {J.}~\bibnamefont {Pugh}},
  \bibinfo {author} {\bibfnamefont {Y.}~\bibnamefont {Gong}}, \bibinfo {author}
  {\bibfnamefont {M.~J.}\ \bibnamefont {Cryan}}, \ and\ \bibinfo {author}
  {\bibfnamefont {T.}~\bibnamefont {Wang}},\ }\bibfield  {title} {\enquote
  {\bibinfo {title} {Monolithically multi-color lasing from an {I}n{G}a{N}
  microdisk on a {S}i substrate},}\ }\href {\doibase
  10.1038/s41598-017-10712-4} {\bibfield  {journal} {\bibinfo  {journal} {Sci.
  Rep.}\ }\textbf {\bibinfo {volume} {7}},\ \bibinfo {pages} {10086} (\bibinfo
  {year} {2017})}\BibitemShut {NoStop}%
\bibitem [{\citenamefont {Mexis}\ \emph {et~al.}(2011)\citenamefont {Mexis},
  \citenamefont {Sergent}, \citenamefont {Guillet}, \citenamefont {Brimont},
  \citenamefont {Bretagnon}, \citenamefont {Gil}, \citenamefont {Semond},
  \citenamefont {Leroux}, \citenamefont {N\'eel}, \citenamefont {David},
  \citenamefont {Checoury},\ and\ \citenamefont {Boucaud}}]{Mexis2011}%
  \BibitemOpen
  \bibfield  {author} {\bibinfo {author} {\bibfnamefont {M.}~\bibnamefont
  {Mexis}}, \bibinfo {author} {\bibfnamefont {S.}~\bibnamefont {Sergent}},
  \bibinfo {author} {\bibfnamefont {T.}~\bibnamefont {Guillet}}, \bibinfo
  {author} {\bibfnamefont {C.}~\bibnamefont {Brimont}}, \bibinfo {author}
  {\bibfnamefont {T.}~\bibnamefont {Bretagnon}}, \bibinfo {author}
  {\bibfnamefont {B.}~\bibnamefont {Gil}}, \bibinfo {author} {\bibfnamefont
  {F.}~\bibnamefont {Semond}}, \bibinfo {author} {\bibfnamefont
  {M.}~\bibnamefont {Leroux}}, \bibinfo {author} {\bibfnamefont
  {D.}~\bibnamefont {N\'eel}}, \bibinfo {author} {\bibfnamefont
  {S.}~\bibnamefont {David}}, \bibinfo {author} {\bibfnamefont
  {X.}~\bibnamefont {Checoury}}, \ and\ \bibinfo {author} {\bibfnamefont
  {P.}~\bibnamefont {Boucaud}},\ }\bibfield  {title} {\enquote {\bibinfo
  {title} {High quality factor nitride-based optical cavities: microdisks with
  embedded {G}a{N}/{A}l({G}a){N} quantum dots},}\ }\href {\doibase
  10.1364/OL.36.002203} {\bibfield  {journal} {\bibinfo  {journal} {Optics
  Letters}\ }\textbf {\bibinfo {volume} {36}},\ \bibinfo {pages} {2203--2205}
  (\bibinfo {year} {2011})}\BibitemShut {NoStop}%
\bibitem [{\citenamefont {Rousseau}\ \emph {et~al.}(2018)\citenamefont
  {Rousseau}, \citenamefont {Callsen}, \citenamefont {Jacopin}, \citenamefont
  {Carlin}, \citenamefont {Butt\'e},\ and\ \citenamefont
  {Grandjean}}]{Rousseau2018}%
  \BibitemOpen
  \bibfield  {author} {\bibinfo {author} {\bibfnamefont {I.}~\bibnamefont
  {Rousseau}}, \bibinfo {author} {\bibfnamefont {G.}~\bibnamefont {Callsen}},
  \bibinfo {author} {\bibfnamefont {G.}~\bibnamefont {Jacopin}}, \bibinfo
  {author} {\bibfnamefont {J.-F.}\ \bibnamefont {Carlin}}, \bibinfo {author}
  {\bibfnamefont {R.}~\bibnamefont {Butt\'e}}, \ and\ \bibinfo {author}
  {\bibfnamefont {N.}~\bibnamefont {Grandjean}},\ }\bibfield  {title} {\enquote
  {\bibinfo {title} {Optical absorption and oxygen passivation of surface
  states in {III}-nitride photonic devices},}\ }\href {\doibase
  10.1063/1.5022150} {\bibfield  {journal} {\bibinfo  {journal} {J. Appl.
  Phys.}\ }\textbf {\bibinfo {volume} {123}},\ \bibinfo {pages} {113103}
  (\bibinfo {year} {2018})}\BibitemShut {NoStop}%
\bibitem [{\citenamefont {Kneissl}\ \emph {et~al.}(2004)\citenamefont
  {Kneissl}, \citenamefont {Teepe}, \citenamefont {Miyashita}, \citenamefont
  {Johnson}, \citenamefont {Chern},\ and\ \citenamefont {Chang}}]{Kneissl2004}%
  \BibitemOpen
  \bibfield  {author} {\bibinfo {author} {\bibfnamefont {M.}~\bibnamefont
  {Kneissl}}, \bibinfo {author} {\bibfnamefont {M.}~\bibnamefont {Teepe}},
  \bibinfo {author} {\bibfnamefont {N.}~\bibnamefont {Miyashita}}, \bibinfo
  {author} {\bibfnamefont {N.~M.}\ \bibnamefont {Johnson}}, \bibinfo {author}
  {\bibfnamefont {G.~D.}\ \bibnamefont {Chern}}, \ and\ \bibinfo {author}
  {\bibfnamefont {R.~K.}\ \bibnamefont {Chang}},\ }\bibfield  {title} {\enquote
  {\bibinfo {title} {Current-injection spiral-shaped microcavity disk laser
  diodes with unidirectional emission},}\ }\href {\doibase 10.1063/1.1691494}
  {\bibfield  {journal} {\bibinfo  {journal} {Appl. Phys. Lett.}\ }\textbf
  {\bibinfo {volume} {84}},\ \bibinfo {pages} {2485--2487} (\bibinfo {year}
  {2004})}\BibitemShut {NoStop}%
\bibitem [{\citenamefont {Feng}\ \emph {et~al.}(2018)\citenamefont {Feng},
  \citenamefont {He}, \citenamefont {Sun}, \citenamefont {Gao}, \citenamefont
  {Li}, \citenamefont {Zhou}, \citenamefont {Liu}, \citenamefont {Zhang},
  \citenamefont {Li}, \citenamefont {Zhang}, \citenamefont {Sun}, \citenamefont
  {Li}, \citenamefont {Wang}, \citenamefont {Ikeda}, \citenamefont {Wang},\
  and\ \citenamefont {Yang}}]{Feng2018}%
  \BibitemOpen
  \bibfield  {author} {\bibinfo {author} {\bibfnamefont {M.}~\bibnamefont
  {Feng}}, \bibinfo {author} {\bibfnamefont {J.}~\bibnamefont {He}}, \bibinfo
  {author} {\bibfnamefont {Q.}~\bibnamefont {Sun}}, \bibinfo {author}
  {\bibfnamefont {H.}~\bibnamefont {Gao}}, \bibinfo {author} {\bibfnamefont
  {Z.}~\bibnamefont {Li}}, \bibinfo {author} {\bibfnamefont {Y.}~\bibnamefont
  {Zhou}}, \bibinfo {author} {\bibfnamefont {J.}~\bibnamefont {Liu}}, \bibinfo
  {author} {\bibfnamefont {S.}~\bibnamefont {Zhang}}, \bibinfo {author}
  {\bibfnamefont {D.}~\bibnamefont {Li}}, \bibinfo {author} {\bibfnamefont
  {L.}~\bibnamefont {Zhang}}, \bibinfo {author} {\bibfnamefont
  {X.}~\bibnamefont {Sun}}, \bibinfo {author} {\bibfnamefont {D.}~\bibnamefont
  {Li}}, \bibinfo {author} {\bibfnamefont {H.}~\bibnamefont {Wang}}, \bibinfo
  {author} {\bibfnamefont {M.}~\bibnamefont {Ikeda}}, \bibinfo {author}
  {\bibfnamefont {R.}~\bibnamefont {Wang}}, \ and\ \bibinfo {author}
  {\bibfnamefont {H.}~\bibnamefont {Yang}},\ }\bibfield  {title} {\enquote
  {\bibinfo {title} {Room-temperature electrically pumped {I}n{G}a{N} based
  microdisk laser grown on {S}i},}\ }\href {\doibase 10.1364/OE.26.005043}
  {\bibfield  {journal} {\bibinfo  {journal} {Opt. Express}\ }\textbf {\bibinfo
  {volume} {26}},\ \bibinfo {pages} {5043--5051} (\bibinfo {year}
  {2018})}\BibitemShut {NoStop}%
\bibitem [{\citenamefont {Wang}\ \emph {et~al.}(2019)\citenamefont {Wang},
  \citenamefont {Feng}, \citenamefont {Zhou}, \citenamefont {Sun},
  \citenamefont {Liu}, \citenamefont {Huang}, \citenamefont {Zhou},
  \citenamefont {Gao}, \citenamefont {Zheng}, \citenamefont {Ikeda},\ and\
  \citenamefont {Yang}}]{Wang2019}%
  \BibitemOpen
  \bibfield  {author} {\bibinfo {author} {\bibfnamefont {J.}~\bibnamefont
  {Wang}}, \bibinfo {author} {\bibfnamefont {M.}~\bibnamefont {Feng}}, \bibinfo
  {author} {\bibfnamefont {R.}~\bibnamefont {Zhou}}, \bibinfo {author}
  {\bibfnamefont {Q.}~\bibnamefont {Sun}}, \bibinfo {author} {\bibfnamefont
  {J.}~\bibnamefont {Liu}}, \bibinfo {author} {\bibfnamefont {Y.}~\bibnamefont
  {Huang}}, \bibinfo {author} {\bibfnamefont {Y.}~\bibnamefont {Zhou}},
  \bibinfo {author} {\bibfnamefont {H.}~\bibnamefont {Gao}}, \bibinfo {author}
  {\bibfnamefont {X.}~\bibnamefont {Zheng}}, \bibinfo {author} {\bibfnamefont
  {M.}~\bibnamefont {Ikeda}}, \ and\ \bibinfo {author} {\bibfnamefont
  {H.}~\bibnamefont {Yang}},\ }\bibfield  {title} {\enquote {\bibinfo {title}
  {Ga{N}-based ultraviolet microdisk laser diode grown on {S}i},}\ }\href@noop
  {} {\bibfield  {journal} {\bibinfo  {journal} {Photonics Research}\ }\textbf
  {\bibinfo {volume} {7}},\ \bibinfo {pages} {B32--B35} (\bibinfo {year}
  {2019})}\BibitemShut {NoStop}%
\bibitem [{\citenamefont {Wang}\ \emph {et~al.}(2020)\citenamefont {Wang},
  \citenamefont {Feng}, \citenamefont {Zhou}, \citenamefont {Sun},
  \citenamefont {Liu}, \citenamefont {Sun}, \citenamefont {Zheng},
  \citenamefont {Ikeda}, \citenamefont {Sheng},\ and\ \citenamefont
  {Yang}}]{Wang2020}%
  \BibitemOpen
  \bibfield  {author} {\bibinfo {author} {\bibfnamefont {J.}~\bibnamefont
  {Wang}}, \bibinfo {author} {\bibfnamefont {M.}~\bibnamefont {Feng}}, \bibinfo
  {author} {\bibfnamefont {R.}~\bibnamefont {Zhou}}, \bibinfo {author}
  {\bibfnamefont {Q.}~\bibnamefont {Sun}}, \bibinfo {author} {\bibfnamefont
  {J.}~\bibnamefont {Liu}}, \bibinfo {author} {\bibfnamefont {X.}~\bibnamefont
  {Sun}}, \bibinfo {author} {\bibfnamefont {X.}~\bibnamefont {Zheng}}, \bibinfo
  {author} {\bibfnamefont {M.}~\bibnamefont {Ikeda}}, \bibinfo {author}
  {\bibfnamefont {X.}~\bibnamefont {Sheng}}, \ and\ \bibinfo {author}
  {\bibfnamefont {H.}~\bibnamefont {Yang}},\ }\bibfield  {title} {\enquote
  {\bibinfo {title} {Continuous-wave electrically injected {G}a{N}-on-{S}i
  microdisk laser diodes},}\ }\href@noop {} {\bibfield  {journal} {\bibinfo
  {journal} {Optics Express}\ }\textbf {\bibinfo {volume} {28}},\ \bibinfo
  {pages} {12201--12208} (\bibinfo {year} {2020})}\BibitemShut {NoStop}%
\bibitem [{\citenamefont {Tabataba-Vakili}\ \emph {et~al.}(2018)\citenamefont
  {Tabataba-Vakili}, \citenamefont {Doyennette}, \citenamefont {Brimont},
  \citenamefont {Guillet}, \citenamefont {Rennesson}, \citenamefont
  {Frayssinet}, \citenamefont {Damilano}, \citenamefont {Duboz}, \citenamefont
  {Semond}, \citenamefont {Roland}, \citenamefont {El$\thinspace$Kurdi},
  \citenamefont {Checoury}, \citenamefont {Sauvage}, \citenamefont {Gayral},\
  and\ \citenamefont {Boucaud}}]{TabatabaVakili2018}%
  \BibitemOpen
  \bibfield  {author} {\bibinfo {author} {\bibfnamefont {F.}~\bibnamefont
  {Tabataba-Vakili}}, \bibinfo {author} {\bibfnamefont {L.}~\bibnamefont
  {Doyennette}}, \bibinfo {author} {\bibfnamefont {C.}~\bibnamefont {Brimont}},
  \bibinfo {author} {\bibfnamefont {T.}~\bibnamefont {Guillet}}, \bibinfo
  {author} {\bibfnamefont {S.}~\bibnamefont {Rennesson}}, \bibinfo {author}
  {\bibfnamefont {E.}~\bibnamefont {Frayssinet}}, \bibinfo {author}
  {\bibfnamefont {B.}~\bibnamefont {Damilano}}, \bibinfo {author}
  {\bibfnamefont {J.-Y.}\ \bibnamefont {Duboz}}, \bibinfo {author}
  {\bibfnamefont {F.}~\bibnamefont {Semond}}, \bibinfo {author} {\bibfnamefont
  {I.}~\bibnamefont {Roland}}, \bibinfo {author} {\bibfnamefont
  {M.}~\bibnamefont {El$\thinspace$Kurdi}}, \bibinfo {author} {\bibfnamefont
  {X.}~\bibnamefont {Checoury}}, \bibinfo {author} {\bibfnamefont
  {S.}~\bibnamefont {Sauvage}}, \bibinfo {author} {\bibfnamefont
  {B.}~\bibnamefont {Gayral}}, \ and\ \bibinfo {author} {\bibfnamefont
  {P.}~\bibnamefont {Boucaud}},\ }\bibfield  {title} {\enquote {\bibinfo
  {title} {Blue microlasers integrated on a photonic platform on silicon},}\
  }\href {\doibase 10.1021/acsphotonics.8b00542} {\bibfield  {journal}
  {\bibinfo  {journal} {ACS Photonics}\ }\textbf {\bibinfo {volume} {5}},\
  \bibinfo {pages} {3643--3648} (\bibinfo {year} {2018})}\BibitemShut {NoStop}%
\bibitem [{\citenamefont {Tabataba-Vakili}\ \emph
  {et~al.}(2019{\natexlab{a}})\citenamefont {Tabataba-Vakili}, \citenamefont
  {Doyennette}, \citenamefont {Brimont}, \citenamefont {Guillet}, \citenamefont
  {Rennesson}, \citenamefont {Damilano}, \citenamefont {Frayssinet},
  \citenamefont {Duboz}, \citenamefont {Checoury}, \citenamefont {Sauvage},
  \citenamefont {El$\thinspace$Kurdi}, \citenamefont {Semond}, \citenamefont
  {Gayral},\ and\ \citenamefont {Boucaud}}]{TabatabaVakili2019_2}%
  \BibitemOpen
  \bibfield  {author} {\bibinfo {author} {\bibfnamefont {F.}~\bibnamefont
  {Tabataba-Vakili}}, \bibinfo {author} {\bibfnamefont {L.}~\bibnamefont
  {Doyennette}}, \bibinfo {author} {\bibfnamefont {C.}~\bibnamefont {Brimont}},
  \bibinfo {author} {\bibfnamefont {T.}~\bibnamefont {Guillet}}, \bibinfo
  {author} {\bibfnamefont {S.}~\bibnamefont {Rennesson}}, \bibinfo {author}
  {\bibfnamefont {B.}~\bibnamefont {Damilano}}, \bibinfo {author}
  {\bibfnamefont {E.}~\bibnamefont {Frayssinet}}, \bibinfo {author}
  {\bibfnamefont {J.-Y.}\ \bibnamefont {Duboz}}, \bibinfo {author}
  {\bibfnamefont {X.}~\bibnamefont {Checoury}}, \bibinfo {author}
  {\bibfnamefont {S.}~\bibnamefont {Sauvage}}, \bibinfo {author} {\bibfnamefont
  {M.}~\bibnamefont {El$\thinspace$Kurdi}}, \bibinfo {author} {\bibfnamefont
  {F.}~\bibnamefont {Semond}}, \bibinfo {author} {\bibfnamefont
  {B.}~\bibnamefont {Gayral}}, \ and\ \bibinfo {author} {\bibfnamefont
  {P.}~\bibnamefont {Boucaud}},\ }\bibfield  {title} {\enquote {\bibinfo
  {title} {Demonstration of critical coupling in an active {III}-nitride
  microdisk photonic circuit on silicon},}\ }\href {\doibase
  10.1038/s41598-019-54416-3} {\bibfield  {journal} {\bibinfo  {journal} {Sci.
  Rep.}\ }\textbf {\bibinfo {volume} {9}},\ \bibinfo {pages} {18095} (\bibinfo
  {year} {2019}{\natexlab{a}})}\BibitemShut {NoStop}%
\bibitem [{\citenamefont {Tabataba-Vakili}\ \emph {et~al.}(2020)\citenamefont
  {Tabataba-Vakili}, \citenamefont {Alloing}, \citenamefont {Damilano},
  \citenamefont {Souissi}, \citenamefont {Brimont}, \citenamefont {Doyennette},
  \citenamefont {Guillet}, \citenamefont {Checoury}, \citenamefont {Kurdi},
  \citenamefont {Chenot}, \citenamefont {Frayssinet}, \citenamefont {Duboz},
  \citenamefont {Semond}, \citenamefont {Gayral},\ and\ \citenamefont
  {Boucaud}}]{TabatabaVakili2020}%
  \BibitemOpen
  \bibfield  {author} {\bibinfo {author} {\bibfnamefont {F.}~\bibnamefont
  {Tabataba-Vakili}}, \bibinfo {author} {\bibfnamefont {B.}~\bibnamefont
  {Alloing}}, \bibinfo {author} {\bibfnamefont {B.}~\bibnamefont {Damilano}},
  \bibinfo {author} {\bibfnamefont {H.}~\bibnamefont {Souissi}}, \bibinfo
  {author} {\bibfnamefont {C.}~\bibnamefont {Brimont}}, \bibinfo {author}
  {\bibfnamefont {L.}~\bibnamefont {Doyennette}}, \bibinfo {author}
  {\bibfnamefont {T.}~\bibnamefont {Guillet}}, \bibinfo {author} {\bibfnamefont
  {X.}~\bibnamefont {Checoury}}, \bibinfo {author} {\bibfnamefont {M.~E.}\
  \bibnamefont {Kurdi}}, \bibinfo {author} {\bibfnamefont {S.}~\bibnamefont
  {Chenot}}, \bibinfo {author} {\bibfnamefont {E.}~\bibnamefont {Frayssinet}},
  \bibinfo {author} {\bibfnamefont {J.-Y.}\ \bibnamefont {Duboz}}, \bibinfo
  {author} {\bibfnamefont {F.}~\bibnamefont {Semond}}, \bibinfo {author}
  {\bibfnamefont {B.}~\bibnamefont {Gayral}}, \ and\ \bibinfo {author}
  {\bibfnamefont {P.}~\bibnamefont {Boucaud}},\ }\bibfield  {title} {\enquote
  {\bibinfo {title} {Monolithic integration of ultraviolet microdisk lasers
  into photonic circuits in a {III}-nitride-on-silicon platform},}\ }\href
  {\doibase 10.1364/OL.395371} {\bibfield  {journal} {\bibinfo  {journal} {Opt.
  Lett.}\ }\textbf {\bibinfo {volume} {45}},\ \bibinfo {pages} {4276--4279}
  (\bibinfo {year} {2020})}\BibitemShut {NoStop}%
\bibitem [{\citenamefont {Tabataba-Vakili}\ \emph
  {et~al.}(2019{\natexlab{b}})\citenamefont {Tabataba-Vakili}, \citenamefont
  {Rennesson}, \citenamefont {Damilano}, \citenamefont {Frayssinet},
  \citenamefont {Duboz}, \citenamefont {Semond}, \citenamefont {Roland},
  \citenamefont {Paulillo}, \citenamefont {Colombelli}, \citenamefont
  {El$\thinspace$Kurdi}, \citenamefont {Checoury}, \citenamefont {Sauvage},
  \citenamefont {Doyennette}, \citenamefont {Brimont}, \citenamefont {Guillet},
  \citenamefont {Gayral},\ and\ \citenamefont
  {Boucaud}}]{TabatabaVakili2019_1}%
  \BibitemOpen
  \bibfield  {author} {\bibinfo {author} {\bibfnamefont {F.}~\bibnamefont
  {Tabataba-Vakili}}, \bibinfo {author} {\bibfnamefont {S.}~\bibnamefont
  {Rennesson}}, \bibinfo {author} {\bibfnamefont {B.}~\bibnamefont {Damilano}},
  \bibinfo {author} {\bibfnamefont {E.}~\bibnamefont {Frayssinet}}, \bibinfo
  {author} {\bibfnamefont {J.-Y.}\ \bibnamefont {Duboz}}, \bibinfo {author}
  {\bibfnamefont {F.}~\bibnamefont {Semond}}, \bibinfo {author} {\bibfnamefont
  {I.}~\bibnamefont {Roland}}, \bibinfo {author} {\bibfnamefont
  {B.}~\bibnamefont {Paulillo}}, \bibinfo {author} {\bibfnamefont
  {R.}~\bibnamefont {Colombelli}}, \bibinfo {author} {\bibfnamefont
  {M.}~\bibnamefont {El$\thinspace$Kurdi}}, \bibinfo {author} {\bibfnamefont
  {X.}~\bibnamefont {Checoury}}, \bibinfo {author} {\bibfnamefont
  {S.}~\bibnamefont {Sauvage}}, \bibinfo {author} {\bibfnamefont
  {L.}~\bibnamefont {Doyennette}}, \bibinfo {author} {\bibfnamefont
  {C.}~\bibnamefont {Brimont}}, \bibinfo {author} {\bibfnamefont
  {T.}~\bibnamefont {Guillet}}, \bibinfo {author} {\bibfnamefont
  {B.}~\bibnamefont {Gayral}}, \ and\ \bibinfo {author} {\bibfnamefont
  {P.}~\bibnamefont {Boucaud}},\ }\bibfield  {title} {\enquote {\bibinfo
  {title} {{III}-nitride on silicon electrically injected microrings for
  nanophotonic circuits},}\ }\href {\doibase 10.1364/OE.27.011800} {\bibfield
  {journal} {\bibinfo  {journal} {Opt. Express}\ }\textbf {\bibinfo {volume}
  {27}},\ \bibinfo {pages} {11800--11808} (\bibinfo {year}
  {2019}{\natexlab{b}})}\BibitemShut {NoStop}%
\bibitem [{\citenamefont {Baba}\ and\ \citenamefont {Sano}(2003)}]{Baba2003}%
  \BibitemOpen
  \bibfield  {author} {\bibinfo {author} {\bibfnamefont {T.}~\bibnamefont
  {Baba}}\ and\ \bibinfo {author} {\bibfnamefont {D.}~\bibnamefont {Sano}},\
  }\bibfield  {title} {\enquote {\bibinfo {title} {Low-threshold lasing and
  {P}urcell effect in microdisk lasers at room temperature},}\ }\href {\doibase
  10.1109/JSTQE.2003.819464} {\bibfield  {journal} {\bibinfo  {journal} {IEEE
  J. Sel. Top. Quantum Electronics}\ }\textbf {\bibinfo {volume} {9}},\
  \bibinfo {pages} {1340--1346} (\bibinfo {year} {2003})}\BibitemShut {NoStop}%
\bibitem [{\citenamefont {Vico$\thinspace$Trivi{\~{n}}o}\ \emph
  {et~al.}(2015)\citenamefont {Vico$\thinspace$Trivi{\~{n}}o}, \citenamefont
  {Butt\'{e}}, \citenamefont {Carlin},\ and\ \citenamefont
  {Grandjean}}]{Trivino2015}%
  \BibitemOpen
  \bibfield  {author} {\bibinfo {author} {\bibfnamefont {N.}~\bibnamefont
  {Vico$\thinspace$Trivi{\~{n}}o}}, \bibinfo {author} {\bibfnamefont
  {R.}~\bibnamefont {Butt\'{e}}}, \bibinfo {author} {\bibfnamefont {J.-F.}\
  \bibnamefont {Carlin}}, \ and\ \bibinfo {author} {\bibfnamefont
  {N.}~\bibnamefont {Grandjean}},\ }\bibfield  {title} {\enquote {\bibinfo
  {title} {Continuous wave blue lasing in {III}-nitride nanobeam cavity on
  silicon},}\ }\href {\doibase 10.1021/nl504432d} {\bibfield  {journal}
  {\bibinfo  {journal} {Nano Lett.}\ }\textbf {\bibinfo {volume} {15}},\
  \bibinfo {pages} {1259--1263} (\bibinfo {year} {2015})}\BibitemShut {NoStop}%
\bibitem [{\citenamefont {Chow}, \citenamefont {Wright},\ and\ \citenamefont
  {Nelson}(1996)}]{Chow1996}%
  \BibitemOpen
  \bibfield  {author} {\bibinfo {author} {\bibfnamefont {W.}~\bibnamefont
  {Chow}}, \bibinfo {author} {\bibfnamefont {A.}~\bibnamefont {Wright}}, \ and\
  \bibinfo {author} {\bibfnamefont {J.}~\bibnamefont {Nelson}},\ }\bibfield
  {title} {\enquote {\bibinfo {title} {Theoretical study of room temperature
  optical gain in {G}a{N} strained quantum wells},}\ }\href@noop {} {\bibfield
  {journal} {\bibinfo  {journal} {Applied physics letters}\ }\textbf {\bibinfo
  {volume} {68}},\ \bibinfo {pages} {296--298} (\bibinfo {year}
  {1996})}\BibitemShut {NoStop}%
\bibitem [{\citenamefont {Hangleiter}\ \emph {et~al.}(1997)\citenamefont
  {Hangleiter}, \citenamefont {Frankowsky}, \citenamefont {H{\"a}rle},\ and\
  \citenamefont {Scholz}}]{Hangleiter1997}%
  \BibitemOpen
  \bibfield  {author} {\bibinfo {author} {\bibfnamefont {A.}~\bibnamefont
  {Hangleiter}}, \bibinfo {author} {\bibfnamefont {G.}~\bibnamefont
  {Frankowsky}}, \bibinfo {author} {\bibfnamefont {V.}~\bibnamefont
  {H{\"a}rle}}, \ and\ \bibinfo {author} {\bibfnamefont {F.}~\bibnamefont
  {Scholz}},\ }\bibfield  {title} {\enquote {\bibinfo {title} {Optical gain in
  the nitrides: are there differences to other {III}--{V} semiconductors?}}\
  }\href@noop {} {\bibfield  {journal} {\bibinfo  {journal} {Materials Science
  and Engineering: B}\ }\textbf {\bibinfo {volume} {43}},\ \bibinfo {pages}
  {201--206} (\bibinfo {year} {1997})}\BibitemShut {NoStop}%
\bibitem [{\citenamefont {Coldren}, \citenamefont {Corzine},\ and\
  \citenamefont {Masanovic}(2012)}]{Coldren2012}%
  \BibitemOpen
  \bibfield  {author} {\bibinfo {author} {\bibfnamefont {L.~A.}\ \bibnamefont
  {Coldren}}, \bibinfo {author} {\bibfnamefont {S.~W.}\ \bibnamefont
  {Corzine}}, \ and\ \bibinfo {author} {\bibfnamefont {M.~L.}\ \bibnamefont
  {Masanovic}},\ }\href@noop {} {\emph {\bibinfo {title} {Diode lasers and
  photonic integrated circuits}}},\ Vol.\ \bibinfo {volume} {218}\ (\bibinfo
  {publisher} {John Wiley \& Sons},\ \bibinfo {address} {Hoboken, New Jersey},\
  \bibinfo {year} {2012})\BibitemShut {NoStop}%
\bibitem [{\citenamefont {Baba}(1997)}]{Baba1997}%
  \BibitemOpen
  \bibfield  {author} {\bibinfo {author} {\bibfnamefont {T.}~\bibnamefont
  {Baba}},\ }\bibfield  {title} {\enquote {\bibinfo {title} {Photonic crystals
  and microdisk cavities based on {GaInAsP-InP} system},}\ }\href@noop {}
  {\bibfield  {journal} {\bibinfo  {journal} {IEEE journal of selected topics
  in quantum electronics}\ }\textbf {\bibinfo {volume} {3}},\ \bibinfo {pages}
  {808--830} (\bibinfo {year} {1997})}\BibitemShut {NoStop}%
\bibitem [{\citenamefont {Van~Exter}, \citenamefont {Nienhuis},\ and\
  \citenamefont {Woerdman}(1996)}]{vanExter1996}%
  \BibitemOpen
  \bibfield  {author} {\bibinfo {author} {\bibfnamefont {M.}~\bibnamefont
  {Van~Exter}}, \bibinfo {author} {\bibfnamefont {G.}~\bibnamefont {Nienhuis}},
  \ and\ \bibinfo {author} {\bibfnamefont {J.}~\bibnamefont {Woerdman}},\
  }\bibfield  {title} {\enquote {\bibinfo {title} {Two simple expressions for
  the spontaneous emission factor $\beta$},}\ }\href@noop {} {\bibfield
  {journal} {\bibinfo  {journal} {Physical Review A}\ }\textbf {\bibinfo
  {volume} {54}},\ \bibinfo {pages} {3553--3558} (\bibinfo {year}
  {1996})}\BibitemShut {NoStop}%
\bibitem [{\citenamefont {Morkoç}(2009)}]{Morkoc2009}%
  \BibitemOpen
  \bibfield  {author} {\bibinfo {author} {\bibfnamefont {H.}~\bibnamefont
  {Morkoç}},\ }\href@noop {} {\emph {\bibinfo {title} {Handbook of Nitride
  Semiconductors and Devices: {G}a{N}-based optical and electronic devices}}}\
  (\bibinfo  {publisher} {Wiley-VCH},\ \bibinfo {address} {Weinheim},\ \bibinfo
  {year} {2009})\BibitemShut {NoStop}%
\bibitem [{\citenamefont {Kawashima}\ \emph {et~al.}(1997)\citenamefont
  {Kawashima}, \citenamefont {Yoshikawa}, \citenamefont {Adachi}, \citenamefont
  {Fuke},\ and\ \citenamefont {Ohtsuka}}]{Kawashima1997}%
  \BibitemOpen
  \bibfield  {author} {\bibinfo {author} {\bibfnamefont {T.}~\bibnamefont
  {Kawashima}}, \bibinfo {author} {\bibfnamefont {H.}~\bibnamefont
  {Yoshikawa}}, \bibinfo {author} {\bibfnamefont {S.}~\bibnamefont {Adachi}},
  \bibinfo {author} {\bibfnamefont {S.}~\bibnamefont {Fuke}}, \ and\ \bibinfo
  {author} {\bibfnamefont {K.}~\bibnamefont {Ohtsuka}},\ }\bibfield  {title}
  {\enquote {\bibinfo {title} {Optical properties of hexagonal {G}a{N}},}\
  }\href@noop {} {\bibfield  {journal} {\bibinfo  {journal} {Journal of Applied
  Physics}\ }\textbf {\bibinfo {volume} {82}},\ \bibinfo {pages} {3528--3535}
  (\bibinfo {year} {1997})}\BibitemShut {NoStop}%
\bibitem [{\citenamefont {Scheibenzuber}\ \emph {et~al.}(2011)\citenamefont
  {Scheibenzuber}, \citenamefont {Schwarz}, \citenamefont {Sulmoni},
  \citenamefont {Dorsaz}, \citenamefont {Carlin},\ and\ \citenamefont
  {Grandjean}}]{Scheibenzuber2011}%
  \BibitemOpen
  \bibfield  {author} {\bibinfo {author} {\bibfnamefont {W.}~\bibnamefont
  {Scheibenzuber}}, \bibinfo {author} {\bibfnamefont {U.}~\bibnamefont
  {Schwarz}}, \bibinfo {author} {\bibfnamefont {L.}~\bibnamefont {Sulmoni}},
  \bibinfo {author} {\bibfnamefont {J.}~\bibnamefont {Dorsaz}}, \bibinfo
  {author} {\bibfnamefont {J.-F.}\ \bibnamefont {Carlin}}, \ and\ \bibinfo
  {author} {\bibfnamefont {N.}~\bibnamefont {Grandjean}},\ }\bibfield  {title}
  {\enquote {\bibinfo {title} {Recombination coefficients of {G}a{N}-based
  laser diodes},}\ }\href@noop {} {\bibfield  {journal} {\bibinfo  {journal}
  {Journal of Applied Physics}\ }\textbf {\bibinfo {volume} {109}},\ \bibinfo
  {pages} {093106} (\bibinfo {year} {2011})}\BibitemShut {NoStop}%
\bibitem [{\citenamefont {Espenlaub}\ \emph {et~al.}(2019)\citenamefont
  {Espenlaub}, \citenamefont {Myers}, \citenamefont {Young}, \citenamefont
  {Marcinkevi{\v{c}}ius}, \citenamefont {Weisbuch},\ and\ \citenamefont
  {Speck}}]{Espenlaub2019}%
  \BibitemOpen
  \bibfield  {author} {\bibinfo {author} {\bibfnamefont {A.~C.}\ \bibnamefont
  {Espenlaub}}, \bibinfo {author} {\bibfnamefont {D.~J.}\ \bibnamefont
  {Myers}}, \bibinfo {author} {\bibfnamefont {E.~C.}\ \bibnamefont {Young}},
  \bibinfo {author} {\bibfnamefont {S.}~\bibnamefont {Marcinkevi{\v{c}}ius}},
  \bibinfo {author} {\bibfnamefont {C.}~\bibnamefont {Weisbuch}}, \ and\
  \bibinfo {author} {\bibfnamefont {J.~S.}\ \bibnamefont {Speck}},\ }\bibfield
  {title} {\enquote {\bibinfo {title} {Evidence of trap-assisted {A}uger
  recombination in low radiative efficiency {MBE}-grown {III}-nitride
  {LED}s},}\ }\href@noop {} {\bibfield  {journal} {\bibinfo  {journal} {Journal
  of Applied Physics}\ }\textbf {\bibinfo {volume} {126}},\ \bibinfo {pages}
  {184502} (\bibinfo {year} {2019})}\BibitemShut {NoStop}%
\bibitem [{\citenamefont {Kozaki}\ \emph {et~al.}(2006)\citenamefont {Kozaki},
  \citenamefont {Matsumura}, \citenamefont {Sugimoto}, \citenamefont
  {Nagahama},\ and\ \citenamefont {Mukai}}]{Kozaki2006}%
  \BibitemOpen
  \bibfield  {author} {\bibinfo {author} {\bibfnamefont {T.}~\bibnamefont
  {Kozaki}}, \bibinfo {author} {\bibfnamefont {H.}~\bibnamefont {Matsumura}},
  \bibinfo {author} {\bibfnamefont {Y.}~\bibnamefont {Sugimoto}}, \bibinfo
  {author} {\bibfnamefont {S.-i.}\ \bibnamefont {Nagahama}}, \ and\ \bibinfo
  {author} {\bibfnamefont {T.}~\bibnamefont {Mukai}},\ }\bibfield  {title}
  {\enquote {\bibinfo {title} {High-power and wide wavelength range {GaN}-based
  laser diodes},}\ }in\ \href@noop {} {\emph {\bibinfo {booktitle} {Novel
  In-Plane Semiconductor Lasers V}}},\ Vol.\ \bibinfo {volume} {6133}\
  (\bibinfo {organization} {International Society for Optics and Photonics},\
  \bibinfo {year} {2006})\ p.\ \bibinfo {pages} {613306}\BibitemShut {NoStop}%
\bibitem [{\citenamefont {Martens}(2018)}]{Martens2018}%
  \BibitemOpen
  \bibfield  {author} {\bibinfo {author} {\bibfnamefont {M.}~\bibnamefont
  {Martens}},\ }\emph {\bibinfo {title} {Optical gain and modal loss in
  {A}l{G}a{N} based deep {UV} lasers}},\ \href@noop {} {Ph.D. thesis},\
  \bibinfo  {school} {Technische Universität Berlin} (\bibinfo {year}
  {2018})\BibitemShut {NoStop}%
\bibitem [{\citenamefont {Rosencher}\ and\ \citenamefont
  {Vinter}(2004)}]{Rosencher2004}%
  \BibitemOpen
  \bibfield  {author} {\bibinfo {author} {\bibfnamefont {E.}~\bibnamefont
  {Rosencher}}\ and\ \bibinfo {author} {\bibfnamefont {B.}~\bibnamefont
  {Vinter}},\ }\href@noop {} {\emph {\bibinfo {title} {Optoelectronics}}}\
  (\bibinfo  {publisher} {Cambridge University Press},\ \bibinfo {address}
  {Cambridge},\ \bibinfo {year} {2004})\BibitemShut {NoStop}%
\bibitem [{\citenamefont {Piprek}(2010)}]{Piprek2010}%
  \BibitemOpen
  \bibfield  {author} {\bibinfo {author} {\bibfnamefont {J.}~\bibnamefont
  {Piprek}},\ }\bibfield  {title} {\enquote {\bibinfo {title} {Efficiency droop
  in nitride-based light-emitting diodes},}\ }\href@noop {} {\bibfield
  {journal} {\bibinfo  {journal} {physica status solidi (a)}\ }\textbf
  {\bibinfo {volume} {207}},\ \bibinfo {pages} {2217--2225} (\bibinfo {year}
  {2010})}\BibitemShut {NoStop}%
\bibitem [{\citenamefont {Karpov}(2015)}]{Karpov2015}%
  \BibitemOpen
  \bibfield  {author} {\bibinfo {author} {\bibfnamefont {S.}~\bibnamefont
  {Karpov}},\ }\bibfield  {title} {\enquote {\bibinfo {title} {{ABC}-model for
  interpretation of internal quantum efficiency and its droop in {III}-nitride
  {LED}s: a review},}\ }\href@noop {} {\bibfield  {journal} {\bibinfo
  {journal} {Optical and Quantum Electronics}\ }\textbf {\bibinfo {volume}
  {47}},\ \bibinfo {pages} {1293--1303} (\bibinfo {year} {2015})}\BibitemShut
  {NoStop}%
\bibitem [{\citenamefont {David}\ \emph {et~al.}(2019)\citenamefont {David},
  \citenamefont {Young}, \citenamefont {Lund},\ and\ \citenamefont
  {Craven}}]{David2020}%
  \BibitemOpen
  \bibfield  {author} {\bibinfo {author} {\bibfnamefont {A.}~\bibnamefont
  {David}}, \bibinfo {author} {\bibfnamefont {N.~G.}\ \bibnamefont {Young}},
  \bibinfo {author} {\bibfnamefont {C.}~\bibnamefont {Lund}}, \ and\ \bibinfo
  {author} {\bibfnamefont {M.~D.}\ \bibnamefont {Craven}},\ }\bibfield  {title}
  {\enquote {\bibinfo {title} {The physics of recombinations in {III}-nitride
  emitters},}\ }\href@noop {} {\bibfield  {journal} {\bibinfo  {journal} {ECS
  Journal of Solid State Science and Technology}\ }\textbf {\bibinfo {volume}
  {9}},\ \bibinfo {pages} {016021} (\bibinfo {year} {2019})}\BibitemShut
  {NoStop}%
\bibitem [{\citenamefont {Arakawa}\ and\ \citenamefont
  {Sakaki}(1982)}]{Arakawa1982}%
  \BibitemOpen
  \bibfield  {author} {\bibinfo {author} {\bibfnamefont {Y.}~\bibnamefont
  {Arakawa}}\ and\ \bibinfo {author} {\bibfnamefont {H.}~\bibnamefont
  {Sakaki}},\ }\bibfield  {title} {\enquote {\bibinfo {title} {Multidimensional
  quantum well laser and temperature dependence of its threshold current},}\
  }\href@noop {} {\bibfield  {journal} {\bibinfo  {journal} {Applied Physics
  Letters}\ }\textbf {\bibinfo {volume} {40}},\ \bibinfo {pages} {939--941}
  (\bibinfo {year} {1982})}\BibitemShut {NoStop}%
\bibitem [{OMS()}]{OMS}%
  \BibitemOpen
  \href@noop {} {\enquote {\bibinfo {title} {{OMS}: 1-{D} mode solver for
  dielectric multilayer slab waveguides},}\ }\bibinfo {howpublished}
  {\url{https://www.computational-photonics.eu/oms.html}},\ \bibinfo {note}
  {accessed: 2020-02-28}\BibitemShut {NoStop}%
\end{thebibliography}%


\newpage

\begin{center}
  \textbf{\large Supplementary information for \\
  Analysis of low-threshold optically pumped III-nitride microdisk lasers} \\[.2cm]

  Farsane Tabataba-Vakili,$^{1,2}$ Christelle Brimont,$^{3}$ Blandine Alloing,$^4$ Benjamin Damilano,$^4$ Laetitia Doyennette,$^3$ Thierry Guillet,$^3$ Moustafa El Kurdi,$^1$  Sébastien Chenot,$^4$  Virginie Brändli,$^4$ Eric Frayssinet,$^4$  Jean-Yves Duboz,$^4$ Fabrice Semond,$^4$ Bruno Gayral$^2$ and Philippe Boucaud$^{4,*}$   \\[.1cm]
   {\itshape ${}^1$Université Paris-Saclay, CNRS, C2N, 91120, Palaiseau, France.\\
  ${}^2$Univ. Grenoble Alpes, CEA, IRIG-Pheliqs, 38000 Grenoble, France.\\
  ${}^3$L2C, Université de Montpellier, CNRS, 34095 Montpellier, France.\\
  ${}^4$Université Côte d’Azur, CNRS, CRHEA, 06560 Valbonne, France.} \\
  ${}^*$Electronic address: philippe.boucaud@crhea.cnrs.fr\\
\end{center}

\date{\today}

\maketitle

\setcounter{equation}{0}
\setcounter{figure}{0}
\setcounter{table}{0}
\setcounter{page}{1}
\renewcommand{\theequation}{S\arabic{equation}}
\renewcommand{\thefigure}{S\arabic{figure}}

\renewcommand{\thepage}{S\arabic{page}} 
\renewcommand{\thesection}{S\arabic{section}}  
\renewcommand{\thetable}{S\arabic{table}}  
\renewcommand{\thefigure}{S\arabic{figure}}

\section*{Material characterization}

\begin{figure}[htbp]
\centering
\includegraphics[width=0.5\linewidth]{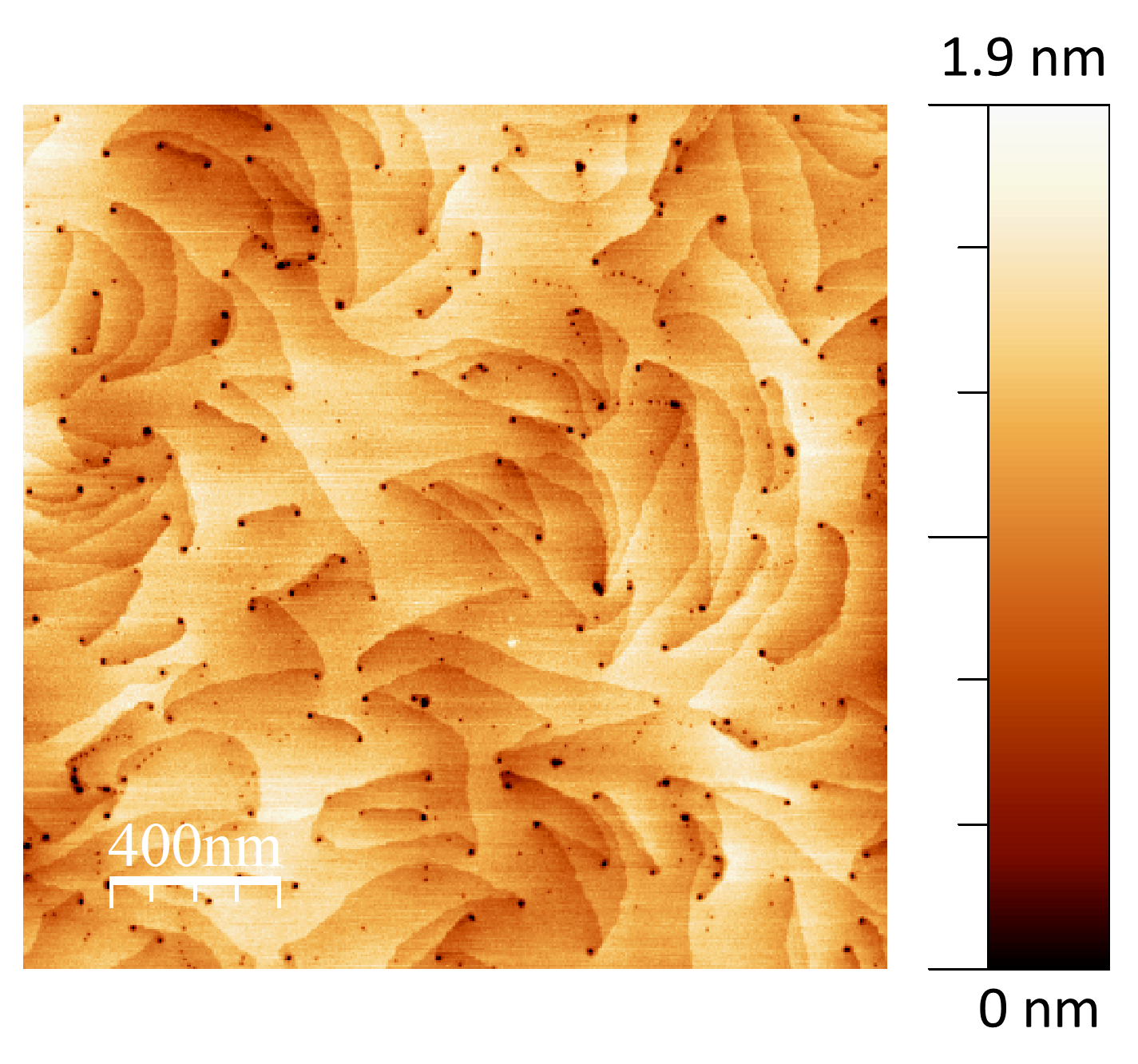}
\caption{$2\times 2$ $\mu \text{m}^2$ AFM image of the sample discussed in the main text.}
\label{fig:afm}
\end{figure}

Fig. \ref{fig:afm} shows a $2\times 2$ $\mu \text{m}^2$ atomic force microscopy (AFM) image of the sample discussed in the main text with a threading dislocation density (TDD) of $1.2\times 10^{10} ~\text{cm}^{-2}$. This is a typical value for thin III-nitride layers on silicon substrate.

\begin{figure}[htbp]
\centering
\includegraphics[width=0.6\linewidth]{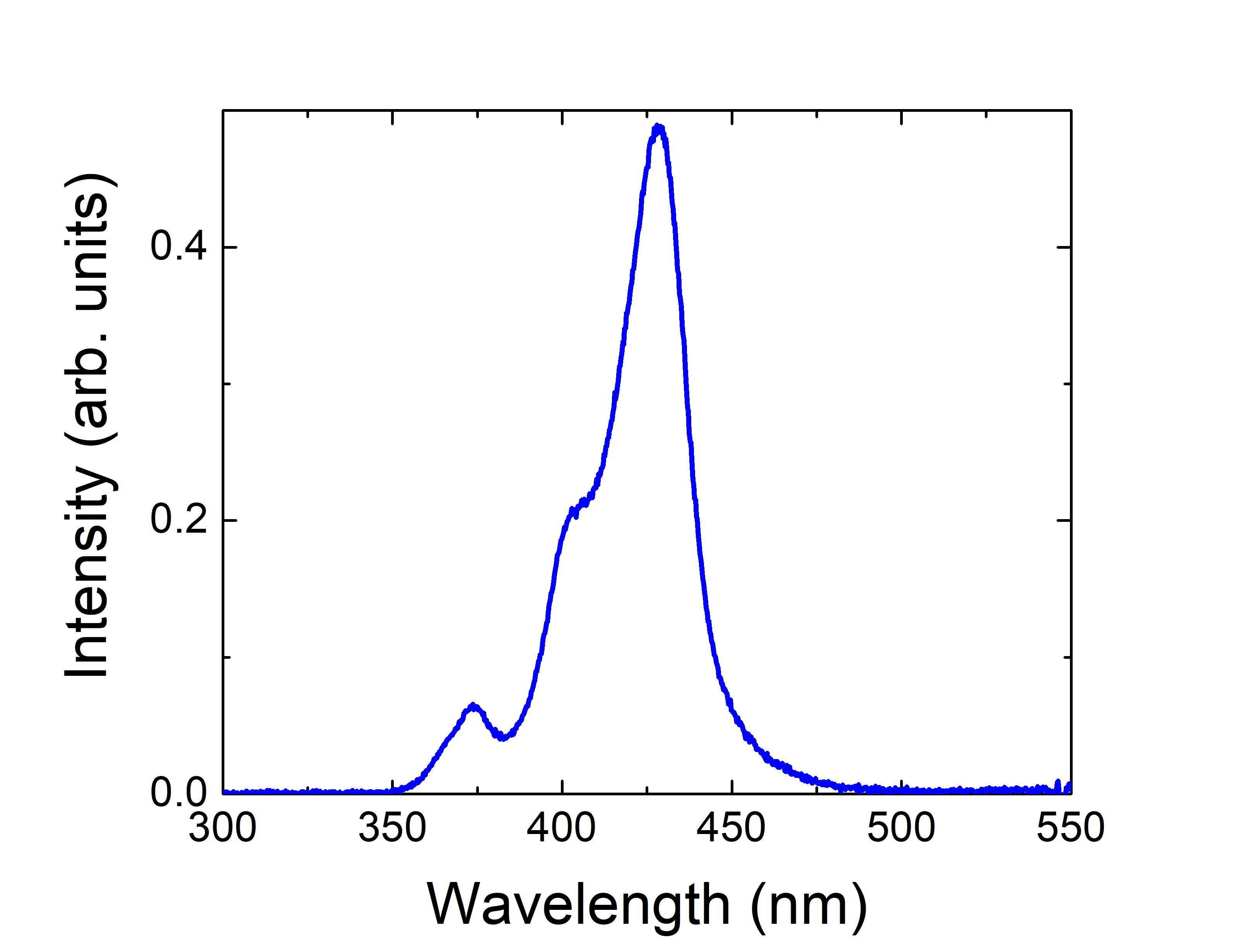}
\caption{High excitation power density $\mu$-PL spectrum of the as-grown sample, measured at $20 ~\text{MW/cm}^2$.}
\label{fig:pl}
\end{figure}

Fig. \ref{fig:pl} shows a room temperature high excitation power density $\mu$-photoluminescence ($\mu$-PL) spectrum of the as-grown sample using a 266 nm pump laser with 400 ps pulse width and 7 kHz repetition rate, measured at $20 ~\text{MW/cm}^2$. The QW emission is centered around 428 nm with a shoulder at 400 nm that is likely related to the emission of excited states. We further observe emission from the tensile strained GaN buffer layer around 373 nm. The full width at half maximum (FWHM) of the main resonance is 22 nm. The main lasing peaks are observed around 423 nm, i.e. on the high-energy side of the  $\mu$-PL spectra.

\section*{Lasing spectra}

\begin{figure}[htbp]
\centering
\includegraphics[width=1\linewidth]{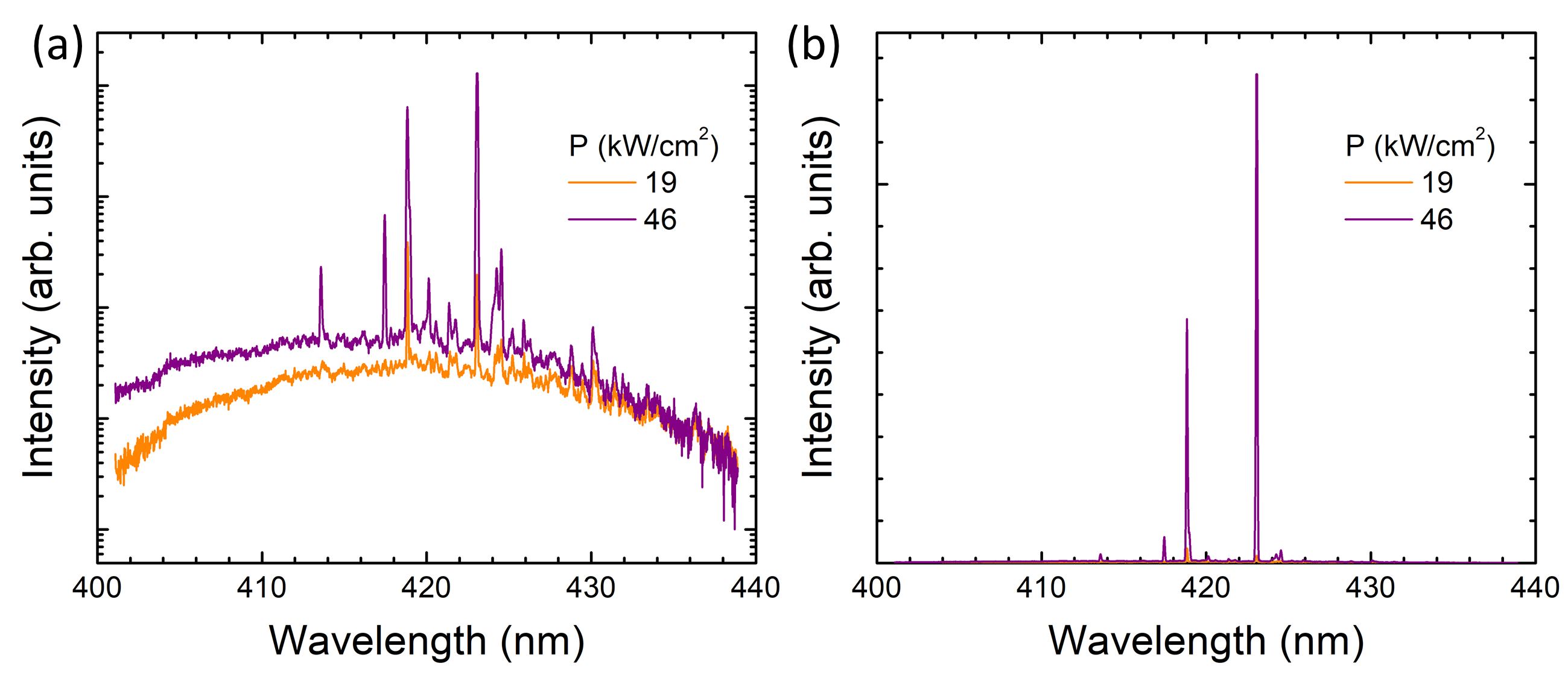}
\caption{Lasing spectra from Fig. 2 at 19 and $46~\text{kW/cm}^2$ with (a) a logarithmic and (b) a linear intensity scale.}
\label{fig:clamp}
\end{figure}

The lasing experiments were performed using a 55 cm focal length spectrometer, with a grating with 1200 lines/mm. The spectral resolution is 0.3 meV, 0.04 nm. The laser spot size was 6 $\mu$m.

Fig. \ref{fig:clamp} shows two lasing spectra of Fig. 2 in the main text at pump powers of 19 and $46~\text{kW/cm}^2$ with (a) a logarithmic and (b) a linear intensity axis to more clearly show the clamping of the background emission in the low energy range (long wavelength). At shorter wavelength the background emission is not fully clamped due to inhomogeneous spectral broadening. The peak to background dynamic is in the range of 300 at $46~\text{kW/cm}^2$, that is $2.6P_{th}$. In Ref. \cite{Tamboli2007}, they only demonstrate a factor of around 8 at $9P_{th}$.

\begin{figure}[htbp]
\centering
\includegraphics[width=0.6\linewidth]{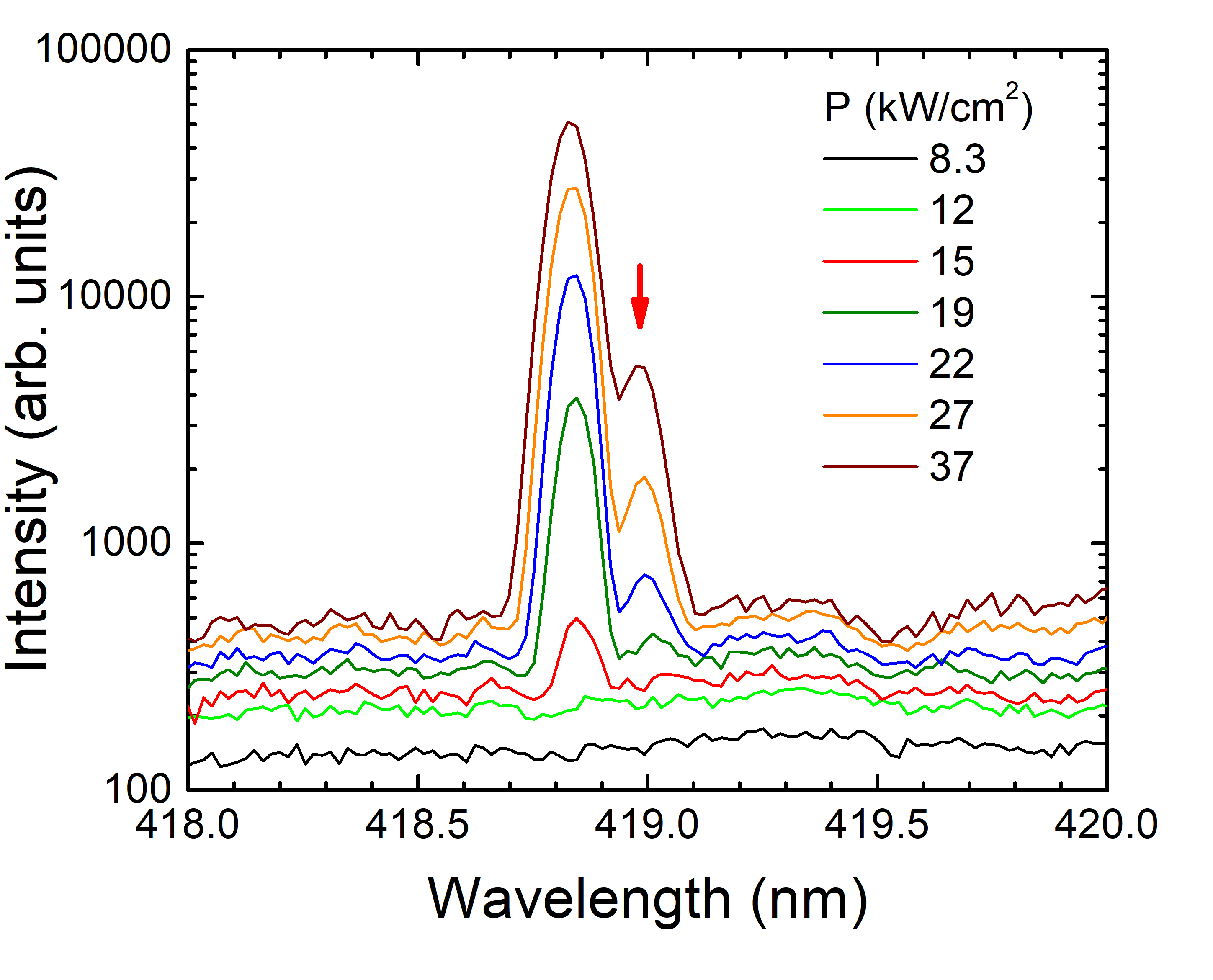}
\caption{Zoom-in of the lasing mode at 419 nm discussed in the main text. The arrow indicates the second lasing mode at nearly the same wavelength.}
\label{fig:419}
\end{figure}

Fig. \ref{fig:419} shows a close-up of the spectra in Fig. 2 around 419 nm. As described in the main text, a second lasing mode appears at higher excitation power at nearly the same wavelength, making the linewidth analysis more complex.

\begin{figure}[htbp]
\centering
\includegraphics[width=0.9\linewidth]{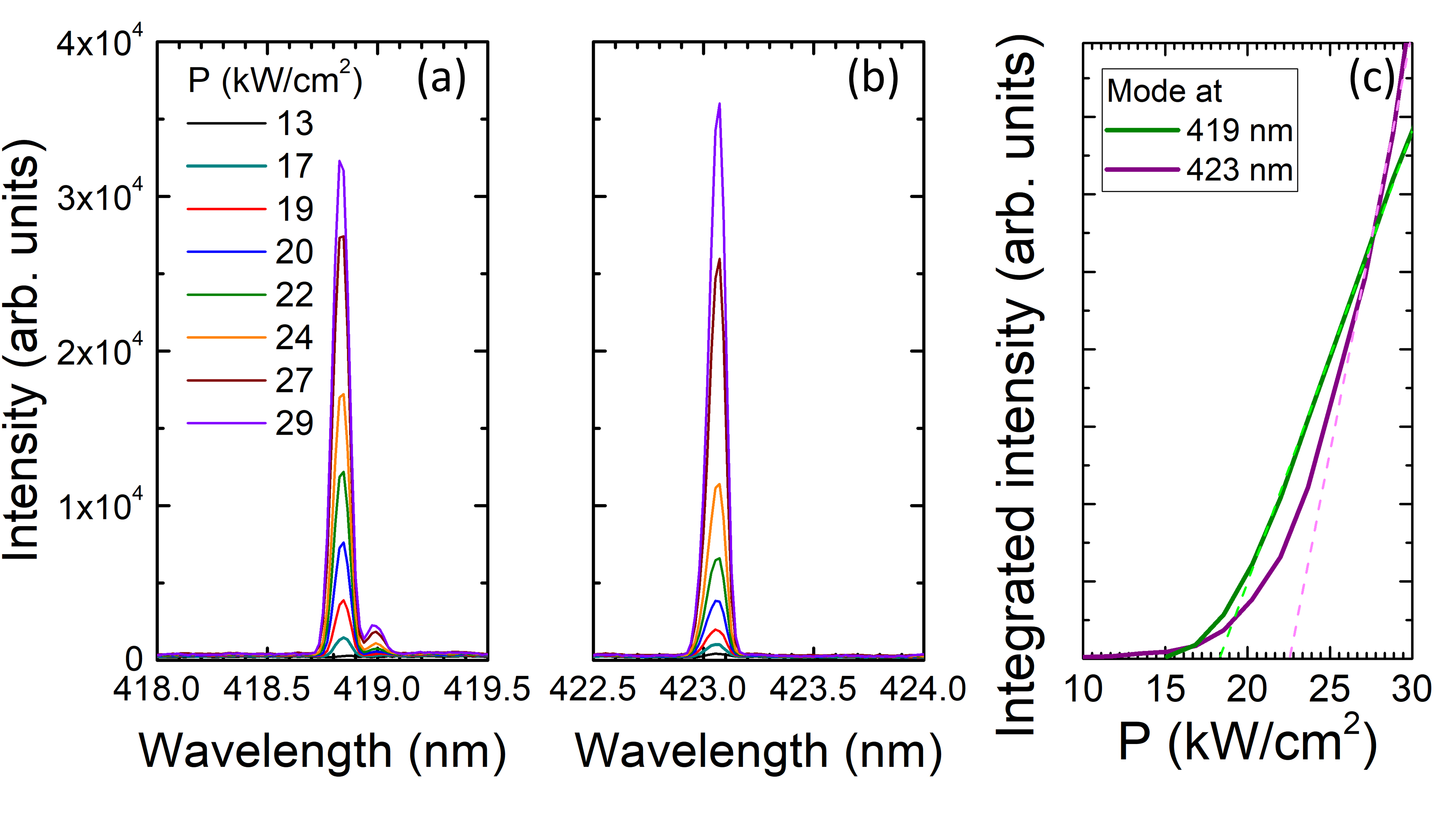}
\caption{Mode competition. Close-up of the modes at (a) 419 nm and (b) 423 nm for pump powers of 13 to 29 kW/cm$^2$. (c) Integrated intensity for both modes with the threshold power density indicated by a straight line.}
\label{fig:comp}
\end{figure}

Figs. \ref{fig:comp} (a) and (b) show the modes at 419 nm and 423 nm, respectively. While initially the mode at 423 nm is more intense, the mode at 419 nm becomes stronger at around 17 kW/cm$^2$. Then at around 29 kW/cm$^2$, the mode at 423 nm takes over again. This mode competiton can also clearly be seen in Fig. \ref{fig:comp} (c), which shows the integrated intensity over pump power for both modes.

\section*{Mode overlap}

\begin{figure}[htbp]
\centering
\includegraphics[width=0.6\linewidth]{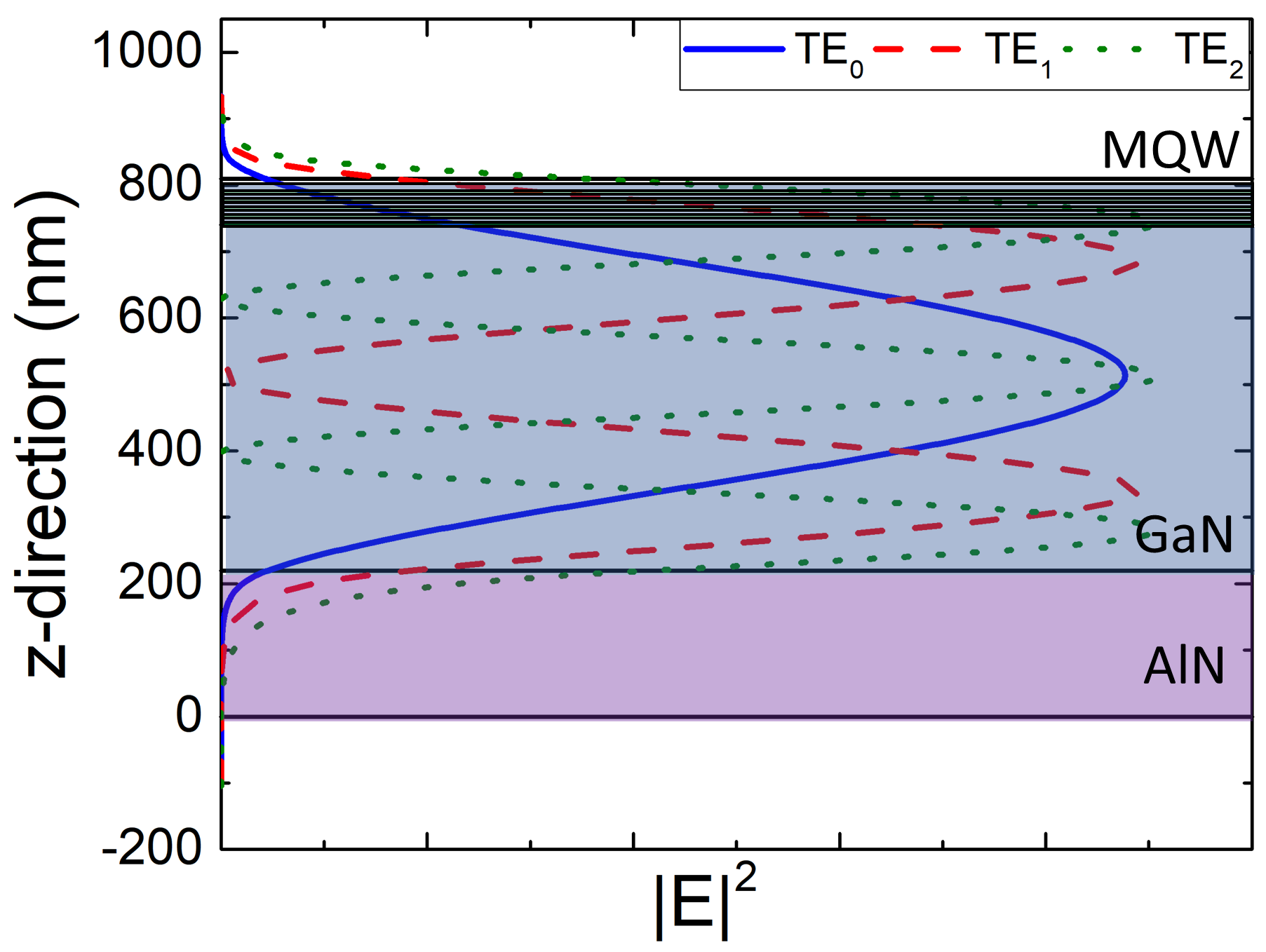}
\caption{Overlap of the TE$_0$, TE$_1$, and TE$_2$ modes with the 5 QWs in a 1D slab waveguide.}
\label{fig:over}
\end{figure}

Figure \ref{fig:over} shows the overlap of the TE$_0$, TE$_1$, and TE$_2$ modes with the 5 QWs in a 1D slab waveguide determined using a 1D mode solver \cite{OMS}. The overlap $\Gamma$ of the 5 QWs with the modes is 0.44\%, 1.6\%, 2.8\% for the TE$_0$, TE$_1$, and TE$_2$ modes, respectively.
Using the same approach, we have calculated an overlap factor of $\Gamma=19.7\%$ for the structure investigated in Ref. \cite{Tamboli2007}, assuming a quantum well width of 4 nm, which was not indicated in their article and is based on information in a previous publication \cite{Haberer2004}.

\section*{Discussion of the threshold gain}

The carrier density at threshold can be described by a different equivalent set of equations. Eq. 2 in the main text can be rewritten as 

\begin{equation}
    n_{th} = n_{tr} \left(1+ \frac{g_{th}}{g_0} \right), \label{eq:nthsupp}
\end{equation}

where $g_{th}$ is the threshold gain that is given by

\begin{equation}
    g_{th} = \frac{n_g}{c \Gamma \tau_c}.\label{eq:gth}
\end{equation}

We assume a constant 2D transparency carrier density of $n_{2D}=3\times 10^{12}~\text{cm}^{-2}$ and an empirical gain coefficient $g_0=35000~\text{cm}^{-1}$ (Ref. \cite{Trivino2015}) for our sample and the results of Refs. \cite{Tamboli2007,Trivino2015}. The 3D transparency carrier density is given by $n_{tr}=n_{2D}/d_{QW}$, where $d_{QW}$ is the thickness of 1 QW.

Using the values $\Gamma=0.44\%$, $n_g=3.44$, $\tau_c=1.3~\text{ps}$ that we give in the main text for our sample, we obtain $g_{th}=20000~\text{cm}^{-1}$. This gives us a factor of $1.57$ between $n_{tr}$ and $n_{th}$.

Meanwhile, for Ref. \cite{Trivino2015}, using $\Gamma=1.7\%$, $n_g=2.81$, and $Q=2600$ that they indicate, we obtain $\tau_c=0.63~\text{ps}$ using Eq. 3 in the main text, and $g_{th}=8700~\text{cm}^{-1}$ and $n_{th}=1.25 n_{tr}$ using Eqs. \ref{eq:nthsupp} and \ref{eq:gth}, which matches well with the values reported in their article.

For Ref. \cite{Tamboli2007}, we obtain $g_{th}=600~\text{cm}^{-1}$ and $n_{th}=1.02 n_{tr}=7.63 \times 10^{18}~\text{cm}^{-3}$ using $\Gamma=19.7\%$, calculated from the structure given in Ref. \cite{Haberer2004}, and $n_g=3$ and $Q=3700$, i.e. $\tau_c =0.84 ~\text{ps}$, as indicated in Ref. 12. A much stronger overlap factor of the quantum wells  leads to a much smaller difference between $n_{th}$ and $n_{tr}$. Note that there is only a factor 2 for the threshold carrier density between our work and Ref. \cite{Tamboli2007}. This is based on the assumption that the QWs in Tamboli et al. are 4 nm thick, which is not explicitly stated. If they are thinner, $n_{tr}$ is larger and $\Gamma$ is smaller.

Furthermore, in a less approximated form, Eq. \ref{eq:nthsupp} can also be written as

\begin{equation}
    n_{th} = n_{tr} e^{\frac{g_{th}}{g_0}}. \label{eq:exp}
\end{equation}

Eq. \ref{eq:exp} gives a ratio of 1.77 between $n_{th}$ and $n_{tr}$, as opposed to 1.57, which was calculated using Eq. \ref{eq:nthsupp}.

\section*{Dependence of the lasing threshold on QW number}

The lasing threshold has a double dependence on the number of QWs, $N_{QW}$. There is a linear dependence that can be seen in Eq. 1 in the main text, as the total thickness of the active region is given by $d=N_{QW} \cdot d_{QW}$. The second non-linear dependence is given in Eq. 2 in the main text, as $\Gamma$ also depends on the QW number and is in first approximation given as $\Gamma=N_{QW} \cdot \Gamma_{QW}$ where $\Gamma_{QW}$ is the overlap for one single quantum well. 

The lasing threshold is thus given by

\begin{equation}
    P_{th} = \frac{E_{phot} N_{QW}d_{QW}}{U \eta_{inj} (1-R)}  \frac{1}{\tau_{tot}}  \left( n_{tr} + \frac{n_g}{c N_{QW}\Gamma_{QW} g \tau_c} \right). \label{eq:Pthsupp}
\end{equation}

\begin{figure}[htbp]
\centering
\includegraphics[width=0.6\linewidth]{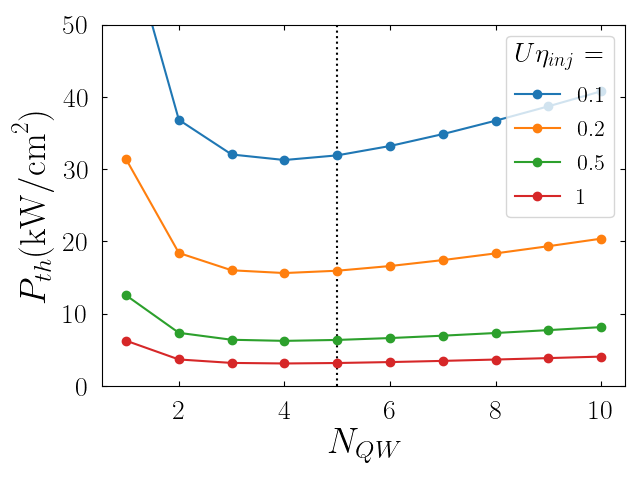}
\caption{$P_{th}$ as a function of $N_{QW}$ using $\Gamma_{QW}=0.088\%$ for different values of $U \eta$.}
\label{fig:PthN}
\end{figure}

Fig. \ref{fig:PthN} shows $P_{th}$ as a function of $N_{QW}$. A minimum can be observed for around 3-4 QWs. However, our 5 QWs are not far from this optimal value for the here considered heterostructure.

\section*{Lasing thresholds for different ABC values}

As discussed in the main text, there is a large variation in the ABC parameters in literature. This variation is obviously a signature of the heterostructure quality and design, including the In content of the layers. We compare the carrier lifetimes and lasing thresholds obtained with the ABC values taken from two articles \cite{Scheibenzuber2011, Espenlaub2019} for our sample as well as the results reported in literature \cite{Tamboli2007, Trivino2015}. Then we further discuss the results of Tamboli et al. \cite{Tamboli2007} using the ABC model discussed by David et al. \cite{David2020}.

\begin{table}[htbp]
\begin{ruledtabular}
\begin{tabular}{@{}lllll@{}}
                   & $U \eta$ & This work & Vico Trivi{\~{n}}o et al. \cite{Trivino2015} & Tamboli et al. \cite{Tamboli2007} \\ \hline
$\tau_{tot}$ (ns)   &          & 5.1       & 6.7            & 11             \\
$P_{th}$ (kW/cm $^2$) & 0.2      & 16        & 2.6            & 5.3            \\
$P_{th}$ (kW/cm $^2$) & 0.5      & 6.4       & 1.0            & 2.1            \\ 
$P_{th}$ (kW/cm $^2$) (experimental) & & 18      & 0.74           & 0.3               \\
\end{tabular}
\caption{Carrier lifetime and threshold power density calculated using the ABC values from Scheibenzuber et al. \cite{Scheibenzuber2011} for our sample as well as Refs. \cite{Trivino2015, Tamboli2007}, as well as the reported experimental values.}
\label{tab:Scheibenzuber}
\end{ruledtabular}
\end{table}

Table \ref{tab:Scheibenzuber} shows $\tau_{tot}$ and $P_{th}$ calculated using the ABC values of Scheibenzuber et al. \cite{Scheibenzuber2011} for our samples as well as for Refs. \cite{Trivino2015, Tamboli2007} for two different values of $U \eta$. The ABC values of Scheibenzuber et al. were obtained around threshold from laser diodes emitting around 415 nm.

Next, we consider the ABC values of Espenlaub et al. (Ref. 35) ($A=5.3 \times 10^8~\text{s}^{-1}$, $B=5 \times 10^{-12}~\text{cm}^3\text{s}^{-1}$, and $C= 6.4 \times 10^{-32}~\text{cm}^6\text{s}^{-1}$), which were obtained for MBE-grown LEDs on sapphire with low efficiency quantum wells. Table \ref{tab:Espenlaub} shows $\tau_{tot}$ and $P_{th}$ calculated using these ABC values for our samples as well as for Refs. \cite{Trivino2015, Tamboli2007} for two different values of $U \eta$.

\begin{table}[htbp]
\begin{ruledtabular}
\begin{tabular}{@{}lllll@{}}
                   & $U \eta$ & This work & Vico Trivi{\~{n}}o et al. \cite{Trivino2015} & Tamboli et al. \cite{Tamboli2007} \\ \hline
$\tau_{tot}$ (ns)   &          & 1.6       & 1.7            & 1.7            \\
$P_{th}$ (kW/cm $^2$) & 0.2      & 50        & 11             & 33             \\
$P_{th}$ (kW/cm $^2$) & 0.5      & 20        & 4.2            & 13             \\ 
$P_{th}$ (kW/cm $^2$) (experimental) & & 18      & 0.74           & 0.3               \\
\end{tabular}
\caption{Carrier lifetime and threshold power density calculated using the ABC values from Espenlaub et al. \cite{Espenlaub2019} for our sample as well as Refs. \cite{Trivino2015, Tamboli2007}, as well as the reported experimental values..}
\label{tab:Espenlaub}
\end{ruledtabular}
\end{table}

Clearly, the calculated thresholds obtained with the ABC values from Scheibenzuber et al. \cite{Scheibenzuber2011} are much closer to the experimental thresholds than the ones calculated using the ABC values of Espenlaub et al. \cite{Espenlaub2019}, which can be easily understood given the higher material quality of the samples investigated in Ref. \cite{Scheibenzuber2011}. We note that the heterostructure discussed in Ref. \cite{Scheibenzuber2011} was grown  on GaN substrate. The same ABC values were used in Ref. \cite{Trivino2015} for cavities with thin III-nitride layers grown on silicon substrates, i.e. in a configuration close to the one that we have investigated in this work.

Lastly, we investigate the dependence of the ABC values on carrier density $n$ and on each other for the structure of Tamboli et al. \cite{Tamboli2007}, based on the review by David et al. \cite{David2020}. Table \ref{tab:ABC_David} shows three sets of ABC parameters. First, we determine $B$ by looking at Fig. 6 (b) of Ref. \cite{David2020} and assuming a QW thickness in the range of 3-4 nm and a threshold carrier density in the range of 0.8 to $1\times 10^{19}~\text{cm}^{-3}$, since the QW thickness is not stated in Ref. \cite{Tamboli2007} (we have been assuming 4 nm until now, based on Ref. \cite{Haberer2004}). Next, we deduce $A$ based on Fig. 7 (a) from Ref. \cite{David2020} and $C$ based on Figs. 9 and 10 (a). 

We can see in Table \ref{tab:ThresholdTamboli} that for the set of parameters labeled David1, the predicted threshold power densities are larger than the one experimentally reported in Ref. \cite{Tamboli2007}. The threshold values are nonetheless closer than those obtained with data from Ref. \cite{Scheibenzuber2011} and Ref. \cite{Espenlaub2019}. Only when the lifetime is in the range of 100 ns and above (David2 and David3) can the theoretical threshold power density reach the 0.3 kW/cm $^2$ range. These values can only be obtained with the highest-quality InGaN quantum wells, as emphasized in Ref.  \cite{David2020}, equivalent to those used in the best LED structures on the market.

\begin{table}[htbp]
\begin{ruledtabular}
\begin{tabular}{@{}llll@{}}

             & David1 & David2 & David3 \\ \hline
$A$ ($10^5~\text{s}^{-1}$)     & 7.9    & 2.5 & 1.6     \\
$B$ ($10^{-13}~\text{cm}^3\text{s}^{-1}$) & 30 &  10   & 7     \\
$C$ ($10^{-32}~\text{cm}^6\text{s}^{-1}$)  & 23 & 4.2   & 2.5 \\
\end{tabular}
\caption{ABC values deduced from Ref. \cite{David2020} starting from three different $B$ values, which depend on QW thickness and carrier density.}
\label{tab:ABC_David}
\end{ruledtabular}
\end{table}

\begin{table}[htbp]
\begin{ruledtabular}
\begin{tabular}{@{}lllll@{}}
                   & $U \eta$ & David1 & David2 & David3 \\ \hline
$\tau_{tot}$ (ns)   &          &  27  & 97 & 140\\
$P_{th}$ (kW/cm $^2$) & 0.2      & 2.2  & 0.6 & 0.41\\
$P_{th}$ (kW/cm $^2$) & 0.5      & 0.87  & 0.24 & 0.16\\ 
\end{tabular}
\caption{Carrier lifetime and threshold power density calculated using three different sets of ABC values deduced from Ref. \cite{David2020} for the results discussed by Tamboli et al. \cite{Tamboli2007}.}
\label{tab:David2}
\label{tab:ThresholdTamboli}
\end{ruledtabular}
\end{table}

\end{document}